\documentclass[%
 reprint,
 amsmath,amssymb,
 aps,
]{revtex4-2}
\usepackage{graphicx}
\usepackage{dcolumn}
\usepackage{bm}
\usepackage[caption=false]{subfig}
\usepackage{url}
\usepackage{float}
\usepackage{subfloat}
\usepackage[export]{adjustbox}
\usepackage{multirow}
\usepackage{color}
\usepackage{fancyhdr}
\usepackage{longtable}
\usepackage{parskip}
\usepackage[T1]{fontenc}
\usepackage[utf8]{inputenc}
\usepackage{multibib}
\usepackage{natbib}
\usepackage{breqn}

\begin{document}
\title{Atlas of Urban Scaling Laws}
\author{Anna Carbone}
\author{Pietro Murialdo}%
\affiliation{%
 Politecnico di Torino Italy 
}%
\author{Alessandra Pieroni}
\affiliation{
Agenzia per l'Italia Digitale \\ Roma Italy
}%
\author{Carina Toxqui-Quitl}
\affiliation{%
 Universidad Politécnica de Tulancingo\\ Hidalgo Mexico
}%

\date{\today}%
\begin{abstract}
Highly accurate  estimates of the urban fractal dimension $D_f$ are obtained by implementing the Detrended Moving Average algorithm (DMA)  on WorldView2 satellite high-resolution  multi-spectral images covering the largest European cities. Higher fractal dimensions are systematically obtained for urban sectors (centrally located areas) than for suburban and peripheral areas, with $Df$ values ranging from  $1.65$ to $1.90$ respectively.The exponents $\beta_s$ and $\beta_i$ of the scaling law $N^{\beta}$ with $N$ the population size, respectively for socio-economic and infrastructural variables, are evaluated for different urban and suburban sectors in terms of the fractal dimension $D_f$.Results confirm the range of empirical values reported in the literature.  Urban scaling laws have been traditionally derived  as if cities were zero-dimensional objects with the relevant feature related to a single homogeneous population value, thus neglecting the microscopic heterogeneity of the urban structure. Our findings allow one to go beyond this limit.  High sensitive and repeatable satellite records yield robust local estimates of the Hurst and scaling exponents. 
Furthermore, the approach allows one to discriminate among different scaling theories, shedding light on the open issue of scaling phenomena, reconciling contradictory scientific perspectives and pave paths to the systematic adoption of the complex system science approach to urban landscape analysis. 
\end{abstract}
\maketitle
\section{Introduction}
\begin{figure}
    \centering
    \includegraphics[width=8cm]{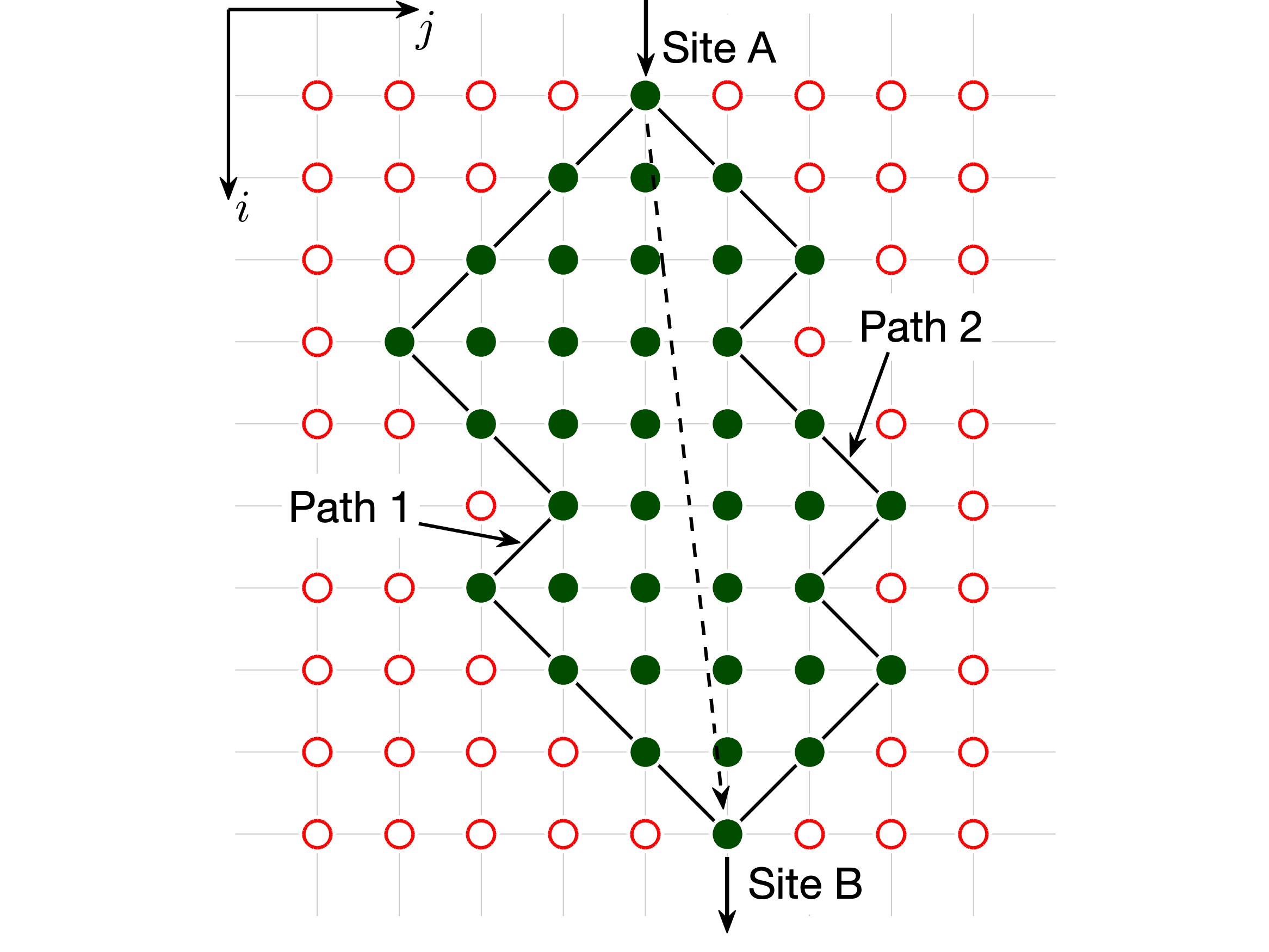}
    \caption{Samples of individual paths between urban sites A and B. Paths have characteristic length $\ell$  and include an area $A$ (cluster area). The length $\ell$  can be related to  the average size (diameter) of the  area $\lambda=A^{1/d}$ and to the fractal dimension $D_f$ by the relationship $\ell \propto \lambda^{D_f}=A^{D_f/d}$.
    }
    \label{fig:SOC}
\end{figure}

The idea of quantifying socio-economic  phenomena in terms of laws derived from  statistical physics and complex systems science continues to spread as highly accurate time and space dependent data become available. Hence, early  studies evidencing  that  diverse socio-economic processes  obey certain empirical laws  can be supported by accurate data analysis and robust statistical modelling \cite{gao2019computational}.
In this context, relevant urban features $Y$ have been linked to the population size $N$ by power-laws with exponents $\beta >1$ typically observed for socio-economic features (e.g. patent production, gross domestic product, crime, pollution), while physical infrastructure features  (e.g. transportation, financial services) tend to increase sublinearly with $\beta<1$ and individual needs (housing, water consumption) with $\beta \approx 1$  (see e.g. \cite{bettencourt2007growth} and references therein).  Despite the diversity of historical and geographical contexts, several aggregated urban features are generally found to scale as power laws of population size, a behaviour whose microscopic origin is still under active debate. 
Diverse theories,   based on dissipative interactions \cite{bettencourt2013origins},  gravity \cite{ribeiro2017model}, three-dimensional fractal buildings \cite{molinero2021geometry}, self-organization \cite{portugali2012self} and synergetics \cite{haken2021urban} have been proposed. Common to these studies is  the dependence of the interactions   on the effective distance $\ell$ connecting any pair of sites, that  for fractal media, can be expressed as $\ell \propto \lambda^{D_f}$  (Fig.~\ref{fig:SOC}).  The exponent $\beta$  is linked to the fractal (Hausdorff) dimension $D_f$ of the background infrastructure,  bridging the  urban scaling  and  fractal geometry  research areas together and thus opening new directions to the quantitative analysis of  socio-economic phenomena and urban complex systems.
\par
Morphology and function  of cities are prominent examples of fractals with the Haudorff dimension providing a measure of the urban concentration across scales  \cite{mandel1,batty1987fractal,frankhauser1998fractal}. 
 The estimation of  fractal dimension in urban contexts begins by analysing the spatial distribution of the build-up area, traditionally performed on cartographic images with black pixels corresponding to build-up space
and  resolution defined by the size of the pixels.
While a uniform distribution of buildings over the investigated area would yield a fractal dimension almost equal to two,  for a detached distribution of buildings,  values lower than two are expected.
 Urban infrastructures cannot be simply quantified  by iteration of  elementary constituents, as it would be appropriate for deterministic fractals,  and require statistically based elaboration of data mapped on the coordinates $i,j$ of the city grid. Methods as diverse as box  and  radial counting  \cite{batty1987fractal,shen2002fractal,frankhauser1998fractal,tannier2013defining,encarnaccao2012fractal,chen2013set},  isarithm, triangular prism, and variogram  \cite{emerson2005comparison,liang2013evaluation,liang2018characterizing} have been adopted to estimate fractal features of  urbanized areas.
Different outcomes have been obtained even for the same city    which might be due to computing-method variations, disparities in image size, map coverage and boundary, image resolution, data accuracy, time period, box-size,
and scale.
\par 
Despite extensive efforts and  several successful applications  \cite{rozenfeld2011area,levinson2012network,yakubo2014superlinear,wu2019transit,keuschnigg2019urban,dong2020understanding}, many issues are still unsolved  preventing full acceptance of the urban scaling ideas \cite{altmann2020spatial,cottineau2017diverse,arcaute2015constructing,rybski2019urban}.   Concerns refer for example  to the microscopic origin of the scaling behaviour and the analytical relationship linking  the exponents $\beta$ and $D_f$, to the proper method and accuracy of statistical fitting. 
The  scaling exponent $\beta$ and the fractal dimension $D_f$ heavily depend on the  definitions, methods and  variables, chosen for their estimation, varying significantly among different works, thus yielding different outcomes and irremediably defying the  intended universality. Heterogeneity and incompleteness of the datasets represent a severe limitation to the accuracy and ultimately prevent  comparability of the  scaling exponents and fractal dimensions across different cities.
High-resolution digitally collected  data have the potential to provide objective definitions and comparable estimates across different regions. In particular, satellite technologies yield  regularly and uniformly recorded data  with well-defined  features, conveniently exploited to gather information about infrastructural and  socio-economic features  \cite{elvidge1997relation,ebener2005wealth,donaldson2016view,jean2016combining,wellmann2020remote,burke2021using}.
However, the ever increasing volume and complexity  of  data pose additional constraints to their practical usability, requiring adequate computational tools. 
\par 
This work addresses some of the above challenges. We provide statistically robust estimates of the  Hurst exponent $H$ and fractal dimensions $D_f$  of urban and suburban sectors  by implementing the high-dimensional \textit{Detrended Moving Average} (DMA)  \cite{carbone2007algorithm} on $1.84m$-resolution Worldview-2 satellite images of several cities   \cite{worldview}. 
For centrally located urban areas  characterized by   regular building grid, fractal dimension values close to $1.9$ are found. Suburban and peripheral areas are characterised by $D_f$ values close to $1.6$. Next, the scaling exponents $\beta_s $ and $\beta_i$ for the socio-economic and infrastructural quantities are estimated. The dependence of  $\beta_s $ and $\beta_i$ on the fractal dimension $D_f$  is discussed on account of the expected behaviour and the empirical values  reported in previous studies. 
The proposed approach, by combining a very accurate computational method and high resolution repeatable satellite records data, yields statistically  robust  estimates of the scaling properties of the urban sector structures. The outcomes are physically sound. 
Overall, the approach could help to discriminate among limited insights and reconcile different controversial scientific perspectives. 
The ultimate goal of this work is the achievement of a shared digital knowledge infrastructure for urban landscape analysis of broad interest. 
\par
The manuscript is organized as follows. In Section \ref{sec:methods} (Definitions and Methods) simple definition of the fractal dimension  and the DMA method  are briefly recalled. In Section \ref{sec:data} (Data and Results) the Worldview-2 satellite images are described and  a few examples are shown and analysed (namely Turin, Wien, Zurich, Prague). The Hurst exponent $H$ and the fractal dimension $D_f$ are estimated for different urban sectors,  compared with previously published results and validated against urban scaling models, in terms of the $\beta$ vs. $D_f$ relationships, in Section \ref{sec:discussion} (Discussion). 
 The main outcomes, potential implications and directions for future work are summarised in Section \ref{sec:conclusions} (Conclusion).
\section{Definitions and Methods}
\label{sec:methods}
Self-similarity concepts and fractal geometry have been extensively adopted to describe real-world random structures characterized by irregular fragmented shapes as well as other complex features that traditional approaches fail to grasp. Generally, scaling relations are obtained for self-similar textures in the form:
\begin{equation}
f(\lambda)\propto   \lambda^{D_f} \quad ,
\label{eq:Nlambda}
\end{equation}
 where  $\lambda$ is the characteristic scale, a measuring unit size, and  $D_f$ the fractal Hausdorff dimension:
\begin{equation}
D_f = d - H \quad ,
\label{eq:Dimension}
\end{equation}
with $d$ the Euclidean embedding dimension and $H$  the Hurst exponent, ranging from $0<H<0.5$ and $0.5<H<1$, respectively  for negatively and positively correlated random sets, and $H=0.5$ corresponding  to the ordinary Brownian function, \textit{i.e.} to fully uncorrelated random sets.
\par
As mentioned in the Introduction,  the high-dimensional detrended moving average ($d$-DMA) \cite{carbone2007algorithm}  is here applied to high-resolution WorldView-2 satellite images  \cite{worldview} to  estimate  the Hurst exponent $H$ and fractal dimension $D_f$ of urban infrastructures.  For the sake of clarity, the main steps of the DMA method are briefly summarized below.
\par 
Random fractal sets can be analytically described in terms of a scalar function $f_H({r}): \mathbb{R}^d\rightarrow \mathbb{R}$   showing self-similarity, with  the Hurst exponent $H$ as a parameter, and correlation depending as a power law on the scale $\lambda$ (a measuring unit size, as in  Eq.~(\ref{eq:Nlambda})). The power-law correlation is reflected by the variance:
\begin{equation}\label{eq:sigmaFBM}
    \sigma_H^2 = \left<[f_H({r}+{\lambda}) - f_H(r)]^2 \right> \propto \parallel{\lambda}\parallel^{2H}
\end{equation}
with ${r} = (x_1,\,x_2,\, ...,\,x_d)$, ${\lambda} = (\lambda_1,\,\lambda_2,\,...,\,\lambda_d)$ and  $\parallel\lambda\parallel = {(\lambda_1^2+\lambda_2^2+...+\lambda_d^2)^{1/2}}$.
\par
The DMA algorithm operates via the definition of a generalized high-dimensional variance $\sigma_{DMA}$  of  $f_H({r})$ around the moving average function $\tilde{f}_H(r)$ \cite{carbone2007algorithm}, that, for $d = 2$, writes:
\begin{dmath}
    \sigma_{DMA}^2 = \frac{1}{(N_1 - n_{1_{max}})(N_2-n_{2_{max}})} 
    \times \sum_{i_1 = n_1}^{N_1} \sum_{i_2 = n_2}^{N_2} [f(i_1,i_2) - \tilde{f}_{n_1 n_2}(i_1,i_2)]^2 \quad ,
    \label{eq:sigmadma1}
\end{dmath}
with $\tilde{f}_{n_1 n_2}(i_1,i_2)$ given by:
\begin{dmath}
    \tilde{f}_{n_1 n_2}(i_1,i_2) = \frac{1}{n_1n_2} \times \sum_{k_1=0}^{n_1-1}\sum_{k_2=0}^{n_2-1} f(i_1-k_1,i_2-k_2) \quad .
\end{dmath}
First, the average scalar field $\tilde{f}_{n_1 n_2}(i_1,i_2)$ is estimated over sub-arrays with different size $n_1\times n_2$. 
The next step of the algorithm is the calculation of the difference $f(i_1,i_2)-\tilde{f}_{n_1,n_2}(i_1,i_2)$ for each sub-array $n_1\times n_2$. It can been shown that Eq. (\ref{eq:sigmadma1}) reduces to the form:
\begin{equation}
    \sigma_{DMA}^2  \sim \left [\sqrt{n_1^2 +n_2^2}\right ]^{2H} = s^{H} \quad ,
       \label{eq:sigmadma2}
\end{equation}
hence a log-log plot of $\sigma_{DMA}^2$ as a function of $s=n_1^2 +n_2^2$ yields a straight line with slope $H$. \par
\begin{figure}[!h]
    \centering
    \includegraphics[width=8cm]{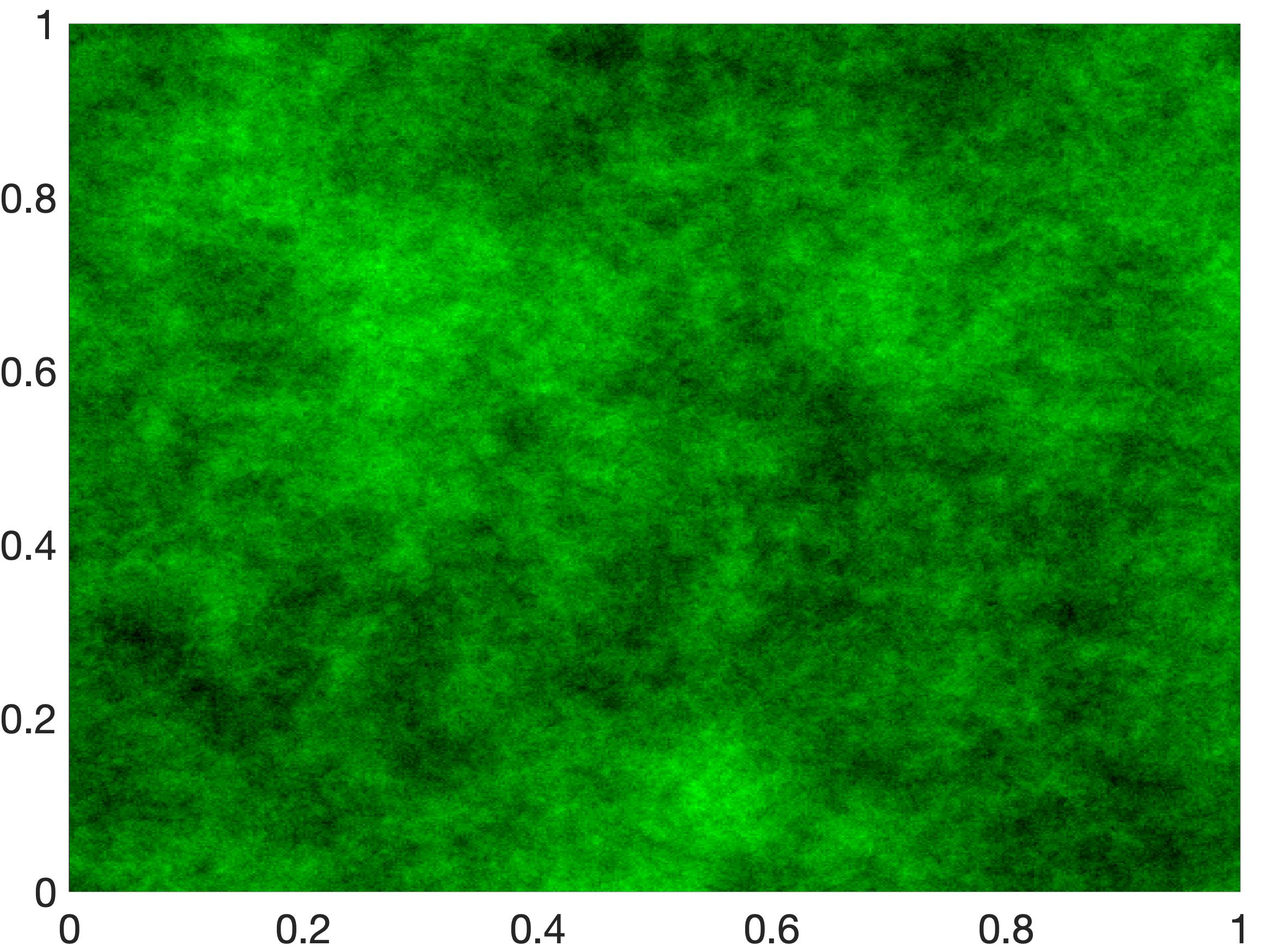}
    \includegraphics[width=8cm]{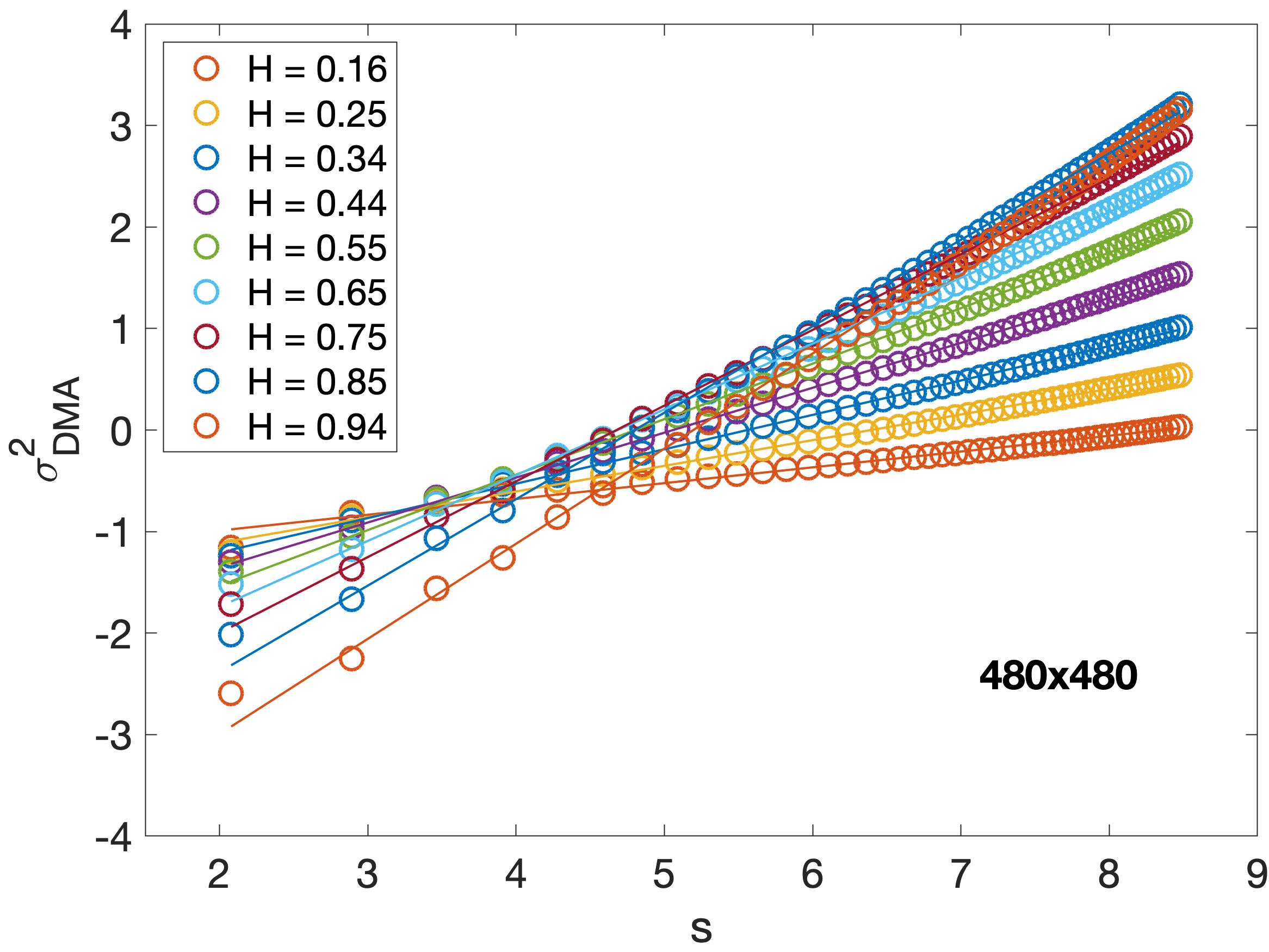}
    \includegraphics[width=8cm]{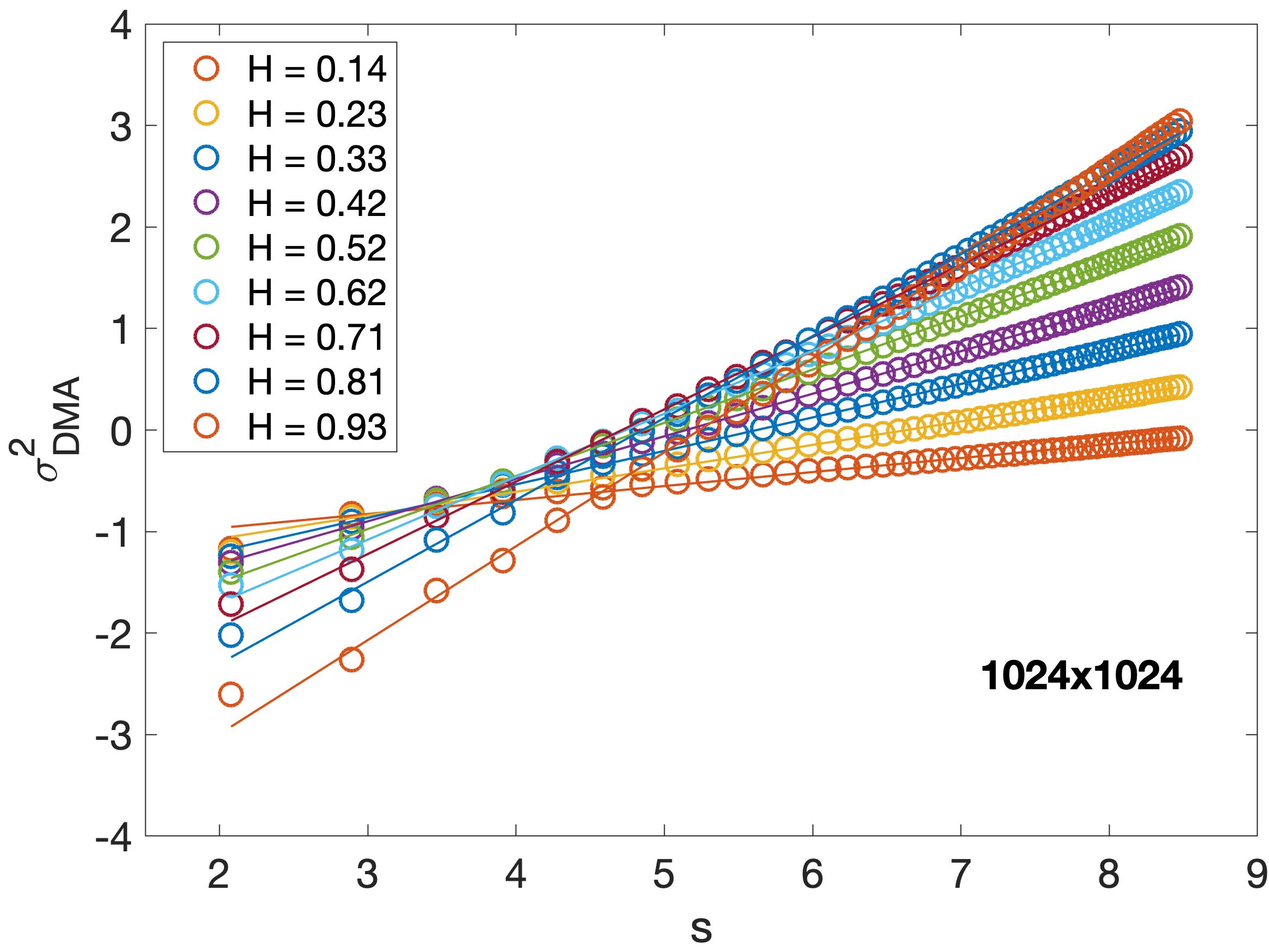}
    \caption{
   Fractional Brownian surface  with size $512\times 512$ and Hurst exponent $H = 0.2$ generated via FRACLAB \citep{FRACLAB} (Top).
 Log-log plots of $\sigma^2_{DMA}$ results for Fractional Brownian surfaces with Hurst exponent $H = 0.1, \, 0.2,\, 0.3, \, ..., 0.9$. 
    Each color refers to $DMA$ results and to Hurst exponent estimates ${H}$ for a different Fractional Brownian surface. 
    The Hurst exponent estimates reported in the legend are obtained as the slope of the regression line by least squares of $\ln{\sigma^2_{DMA}}$.
    Results are reported for Fractional Brownian surfaces of size $480 \times 480$ (middle panel) and $1024 \times 1024$ (bottom panel).}
    \label{fig:fbm}
\end{figure}

 The scaling behaviour expected by Eq.~(\ref{eq:sigmadma2}) is illustrated in Fig.~(\ref{fig:fbm}) where the $2d$-DMA method  is implemented on artificial fractal images, with different  size and Hurst exponent,  generated  by  Cholesky-Levinson Factorization (CLF)  \cite{FRACLAB}. One of such surfaces with  $H=0.2$  is shown in Fig. \ref{fig:fbm} (top panel). The $\sigma_{DMA}$ values obtained for artificial fractal surfaces with input Hurst exponent ranging from $0.1$ to $0.9$, size $480\times480$ and $1024\times1024$ are plotted on log-log scales (middle and bottom panels).
The difference between the input Hurst exponents and the  DMA outcomes is negligible and decreases as the size of the surface increases.
\par
Real-world random data sets  are not ideal fractals, as those defined by the fractional Brownian functions $f_H({r})$, which are defined to exist at all scales.  Being characterized by finite sizes that set  upper and lower limits to the small and large observable scales, deviations from the ideality should be expected. As a rule,  real-world random data sets are classified as \textit{fractals} if their variance can be approximated by a power law over at least three decades of scales.
\par 
The high-dimensional \textit{Detrended Moving Average} (DMA) has been applied to $2d$ and $3d$ artificial structures in \cite{carbone2010snow,turk2010fractal}.
  Evolution of  rural landscape  of Mangystan (Kazakhstan) and  New Mexico (USA) monthly recorded from July 1982 to May 2012 by the multi-spectral  LandSat Thematic Mapper (TM)   have been analysed in \cite{valdiviezo2014hurst}.
Hurst exponents ranging between $0.21 \leq H \leq 0.30$ and  $0.11 \leq H \leq 0.30$, corresponding to  fractal dimensions  between $1.70 \leq D_f \leq 1.79$ and $1.70 \leq D_f \leq 1.89$, have been found respectively for Mangystan and New Mexico. Fractal dimension increases over time as man-made infrastructures and build-up areas grow at the expenses of the natural landscape.  In the next section, the fractal dimension of  WorldView-2 satellite images of several cities will be estimated by using the two-dimensional Detrending Moving Average algorithm (DMA). 
\section{Data and Results}
\label{sec:data}
 WorldView-2  \cite{worldview}  provides panchromatic imagery with $0.46\textrm{m}$ resolution, and eight-band multispectral imagery with $1.84 \textrm{m}$ resolution - representing one of the highest available spaceborne resolutions.  
The subset \textit{European Cities} of the WorldView-2 database includes  images of several European cities and their hinterland, processed by the European Space Imaging GmbH  during February 2011 to October 2013. The collection is related to ESA’s EO missions for the coverage of the urban areas in Europe and is referred as the \textit{Urban Atlas}.  With spatial resolutions of the order of $10m-30m$, LandSat and Sentinel satellites are very effective at mapping land coverage and criosphere  by identifying the spectral signature and broadly classifying areas containing that spectral pattern.  Multi-spectral satellite imagery with pixel resolution  of the order of $1m$ and less    provide  finer scale features able to investigate Earth crust phenomena at a microscopic level. The high resolution  might enable to discriminate fine details of Land Use/Land Cover (LULC) such as farmland, urban areas, quality of road surfaces, and health of plants. The multiple spectral bands  yield inter-band spectral information  to discriminate features of texture  \cite{arreola2021non,safia2015multiband}.
\par
\begin{figure}[!htb]
    \centering
    \includegraphics[width=8cm]{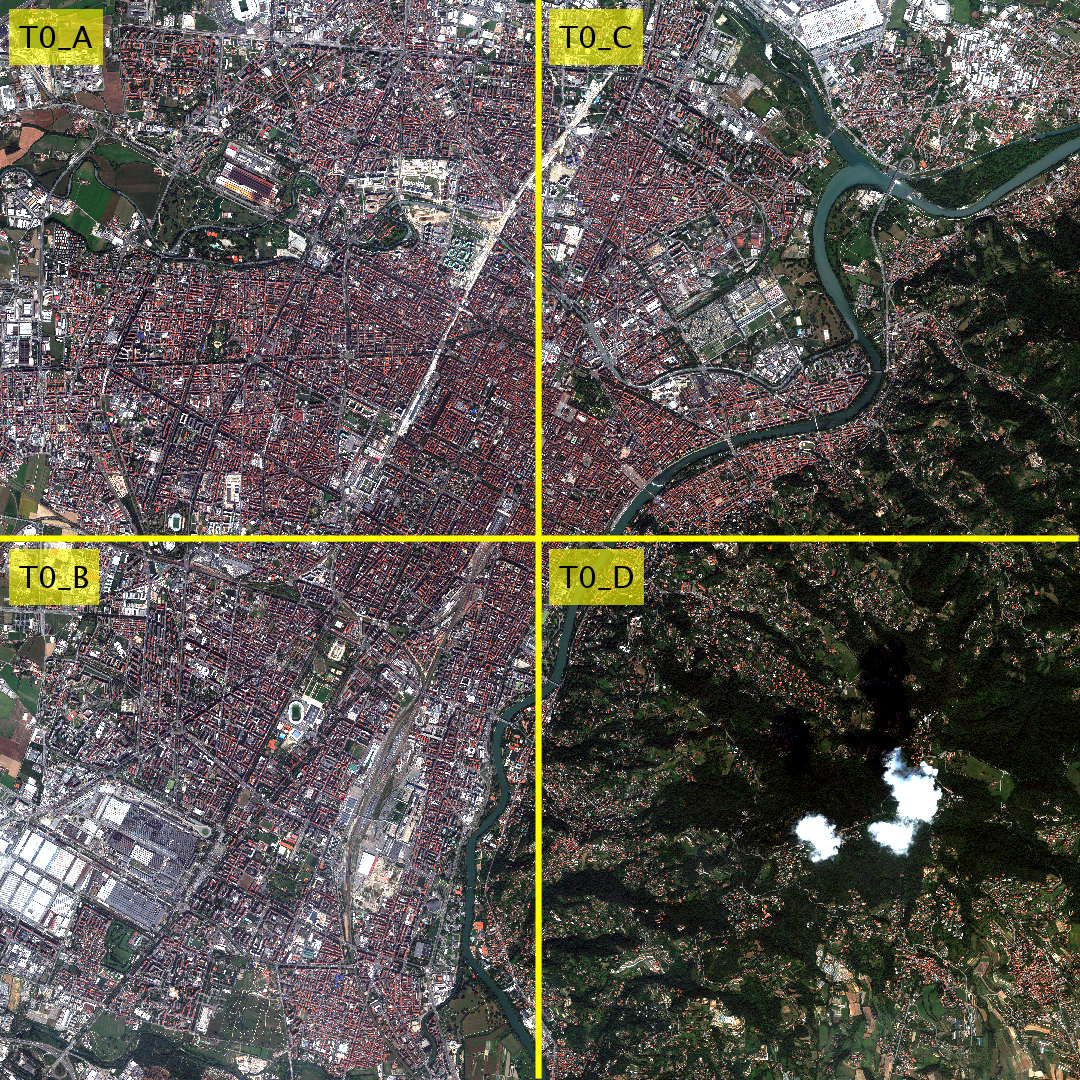}
    \includegraphics[width=8cm]{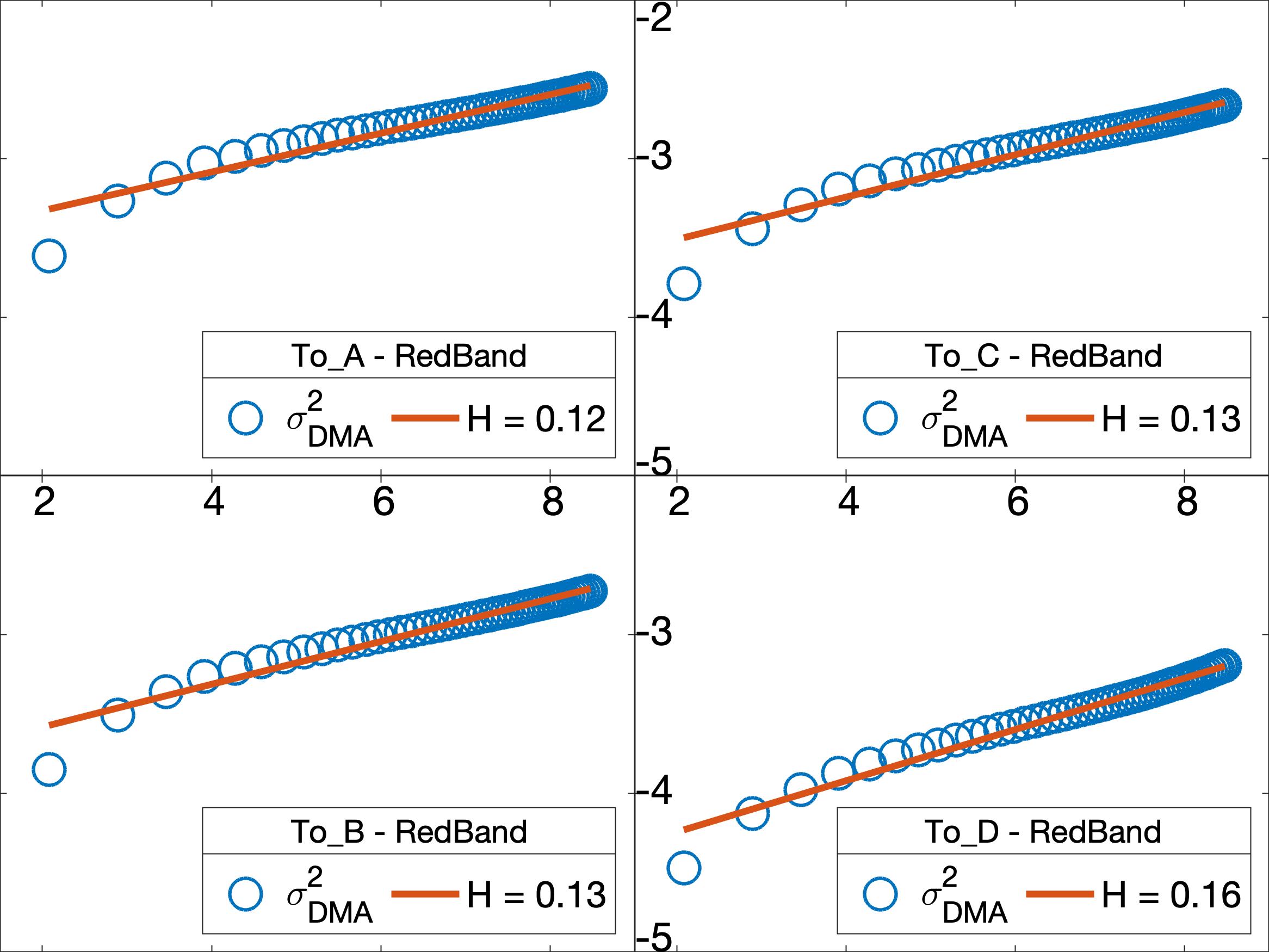}
    \caption{
    Image 45-024 (Torino) downloaded from the \textit{Urban Atlas} collection of the largest European Cities of WorldView-2 satellite images \cite{worldview}. 
    The image is multi-spectral with size $1080 \times 1080$. 
    Yellow lines divide the image into 4 sub-images \textit{A, B, C, D} of size $540 \times 540$  ({Top}).
   Log-log plots of $\sigma^2_{DMA}$ for sub-images \textit{A, B, C, D}. The DMA results refer to the red band only. 
    Hurst Exponent estimates ${H}$, obtained as the slope of the regression line by least squares, are calculated for each sub-image.
    Goodness of fit is evaluated by $R^2$  ({Bottom})
    }
   \label{fig:WV2_TO}
\end{figure}
\begin{figure}[!htb]
    \centering
    \includegraphics[width=8cm]{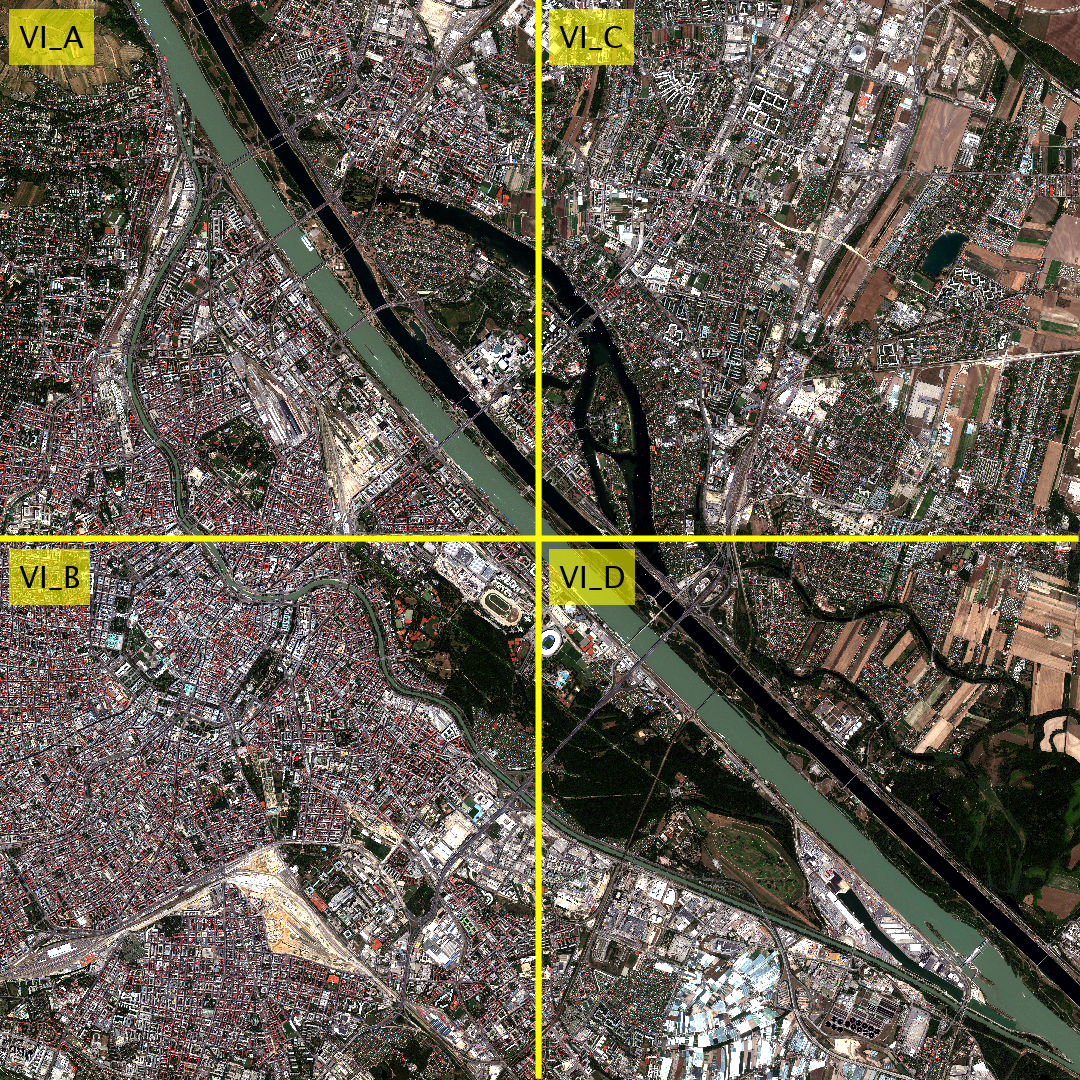}
    \includegraphics[width=8cm]{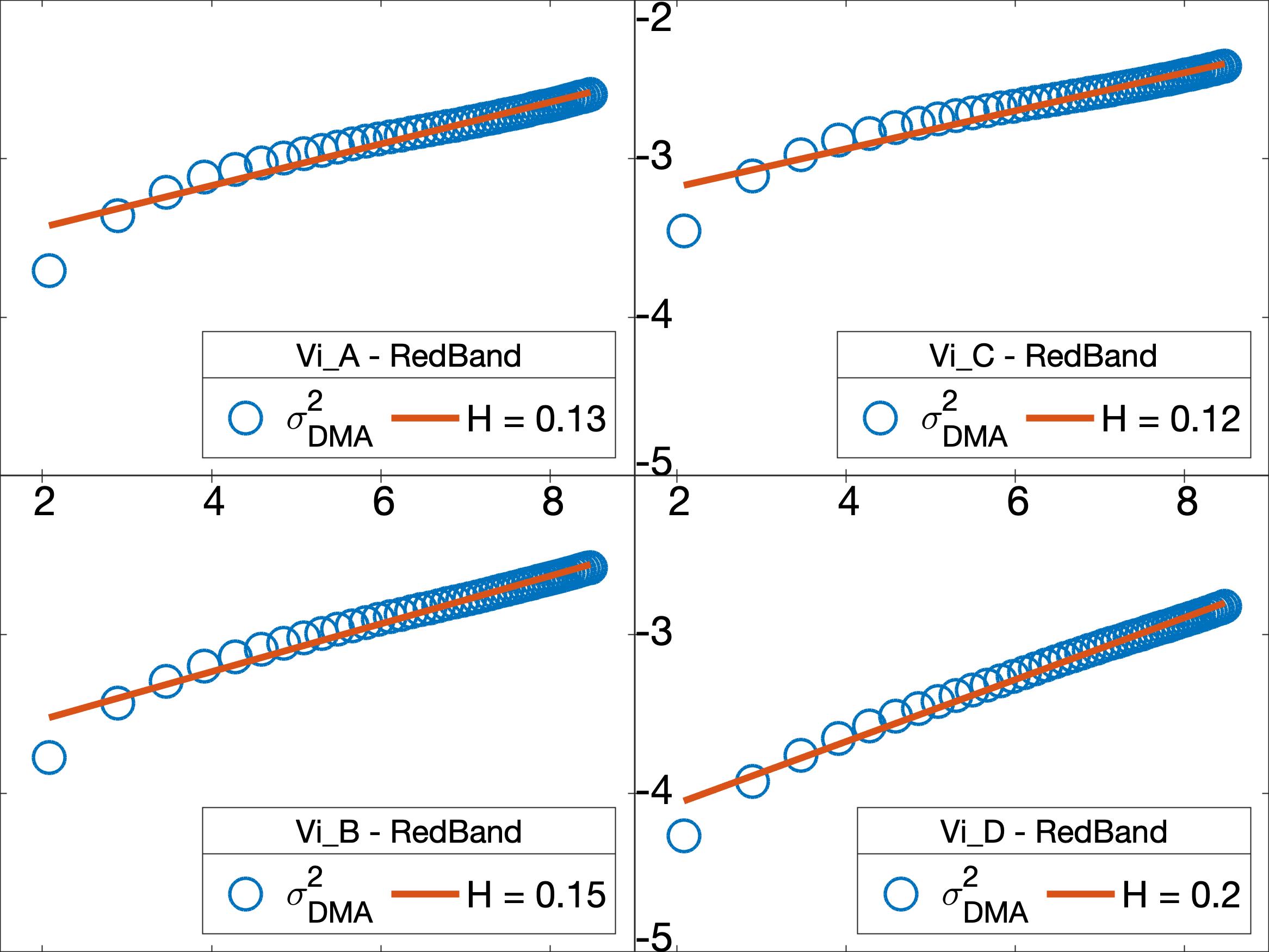}
    \caption{      
    Same as for Fig. \ref{fig:WV2_TO} but for  image 48-181 (Vienna).
    }
    \label{fig:WV2_WI}
\end{figure}
\begin{figure}[htb]
    \centering
    \includegraphics[width=8cm]{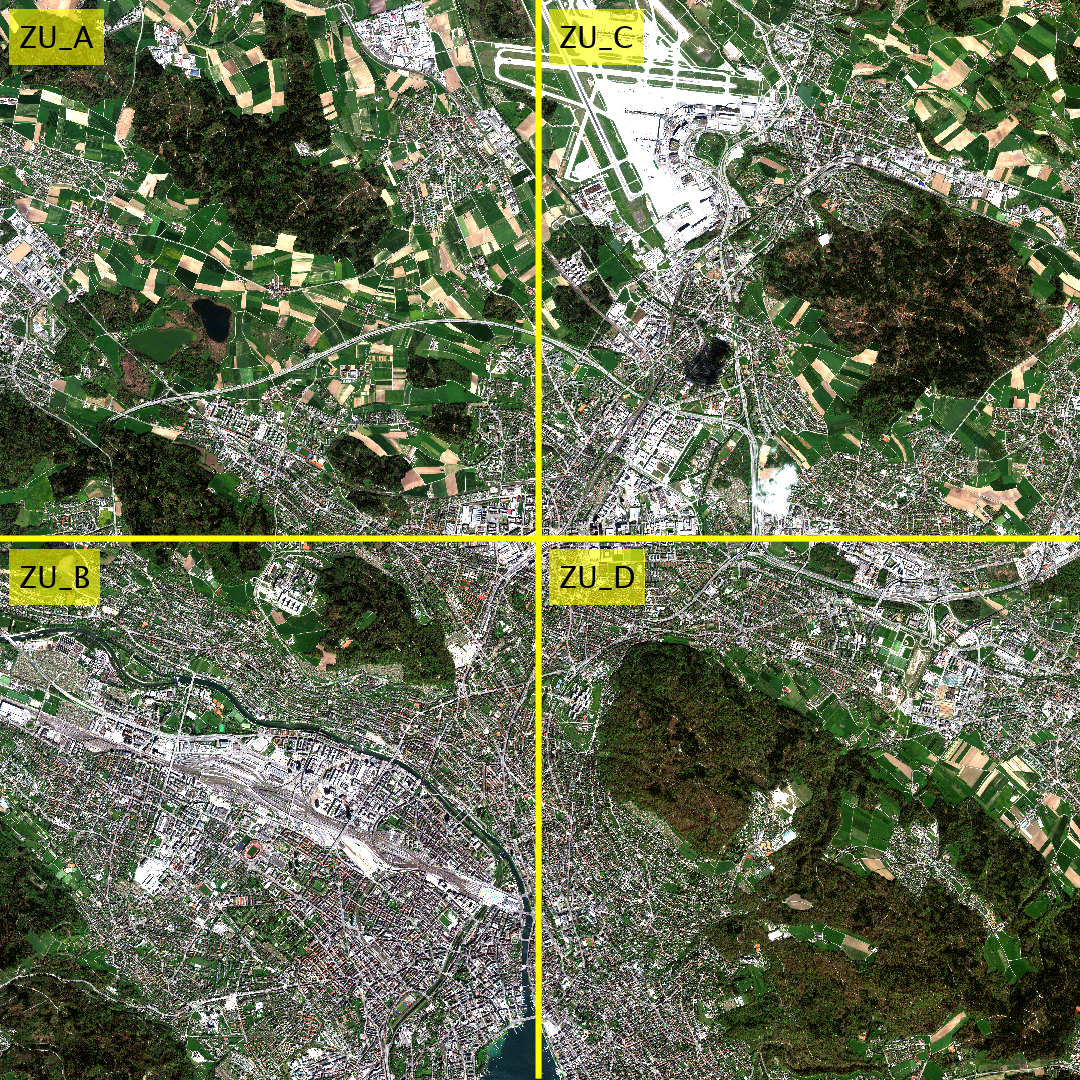}
    \includegraphics[width=8cm]{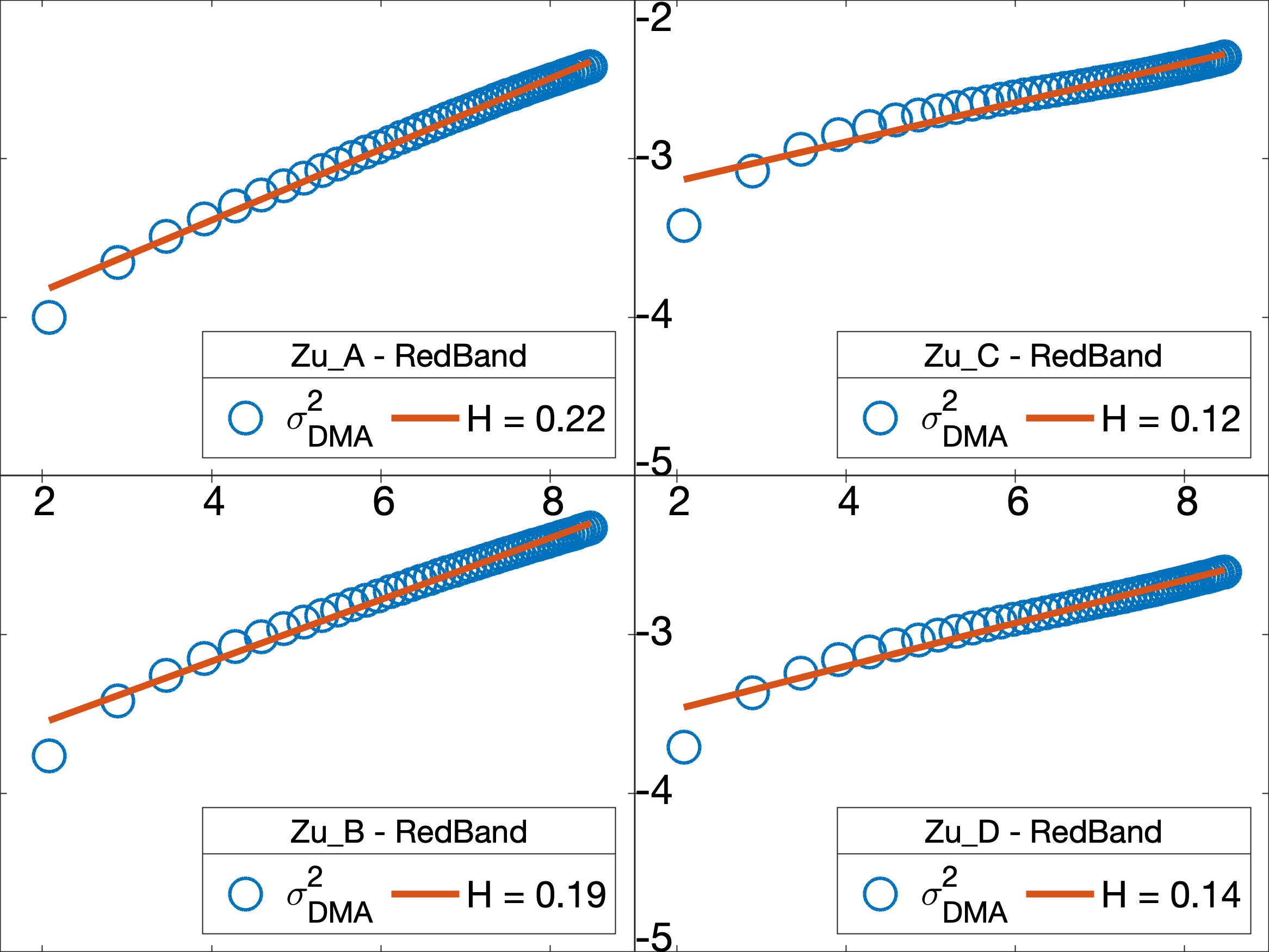}
    \caption{Same as Fig. \ref{fig:WV2_TO}  but for the image  47-377 (Zurich).
    }
    \label{fig:WV2_ZU}
\end{figure}
\begin{figure}[!htb]
    \centering
    \includegraphics[width=8cm]{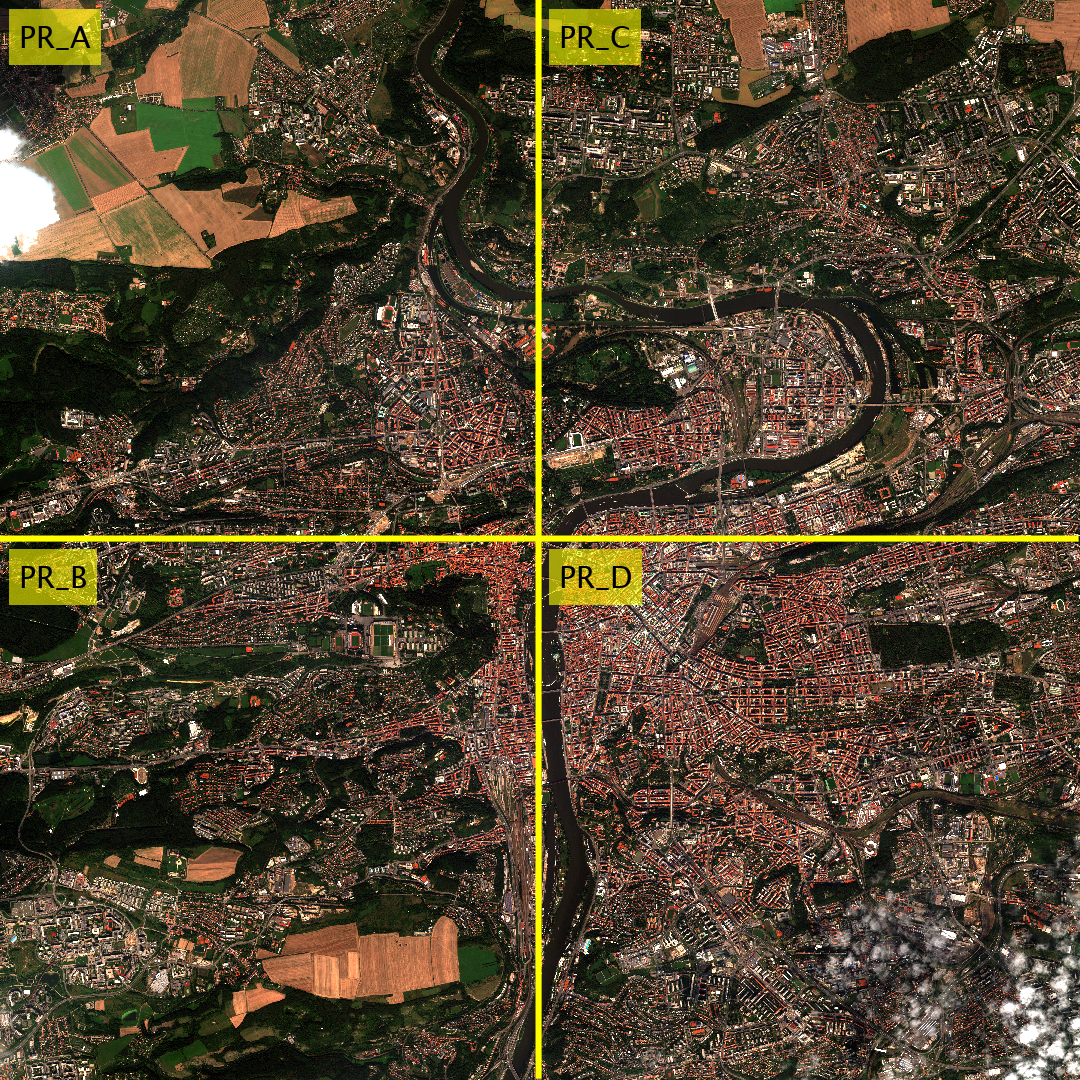}
    \includegraphics[width=8cm]{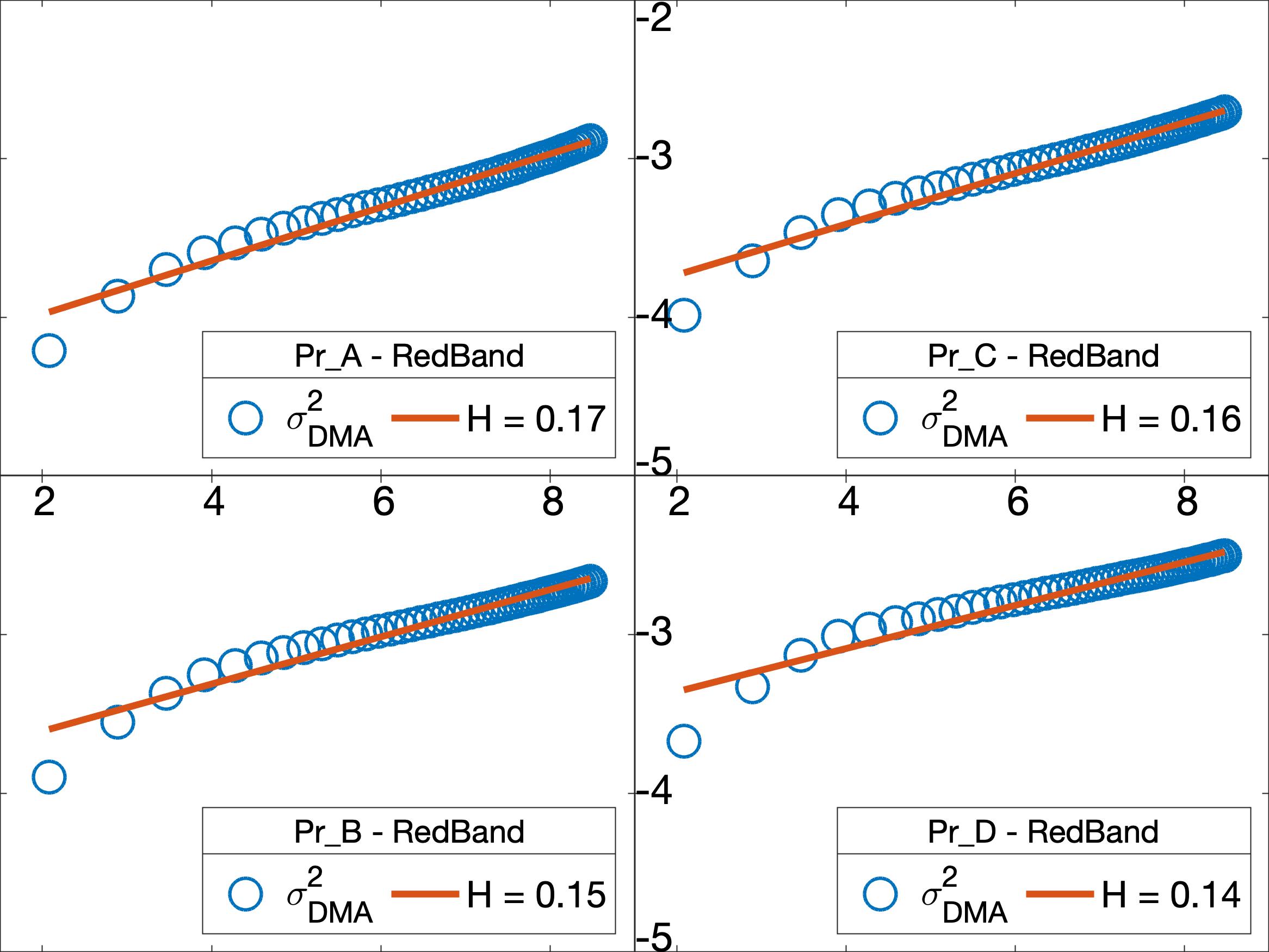}
    \caption{ Same as Fig. \ref{fig:WV2_TO}  but for the image  N50-090  (Prague).
    }
    \label{fig:WV2_PR}
\end{figure}

Samples of the analysed urban areas are shown in the top panels of  Figs. \ref{fig:WV2_TO}-\ref{fig:WV2_PR}. The images  are $1080\times 1080$ pixels large. Sub-images are obtained by  dividing the main image  into four squares of size $540\times 540$, delimited by yellow lines and labelled by A, B, C, D. 
Here, we report results obtained on a single band, i.e. the red band. Results  obtained for  green and blue bands,  different sectors and other cities will be reported in a forthcoming work.
Before implementing the DMA algorithm, raw data  are converted from the \textit{uint8} to the \textit{double} format
For each sub-image, the  algorithm is implemented separately to grasp the variability of the scaling properties of different areas (partially mountainous, suburban, centrally located areas).
\par
Log-log values of the  $\sigma^2_{DMA}$ are plotted in the bottom panels of  Figs. \ref{fig:WV2_TO}-\ref{fig:WV2_PR}. Deviations from the fully linear trend, that would be expected for an ideal fractal, can be  observed particularly at the low scales (small $s$ values) where the $\sigma^2_{DMA}$ drops down. 
In order to account  for  and evaluate the extent of non-ideality and the deviations at the extreme scales,  multiple computational steps are implemented.  
Regressions are computed for the first three decades  ($2<s<5$), the last three decades ($5.5<s<8.5$)  and the  whole range of scales, providing three estimates of the Hurst exponent   (respectively $H_{1}$, $H_{2}$ and ${H}$). The last three decades and the  whole range provide quite close and accurate values of the Hurst exponents (${H}$  and  $H_{2}$) with  excellent goodness of fit as indicated  by the high $R^2$ values. Higher values of the slope are obtained for the first decade ($H_{1}$).  Being the pixel resolution of the order of $1.84 \textrm{m}$ (single band),  the minimum area   detectable by the DMA algorithm is of the order of $1.84\textrm{m} \times 1.84\textrm{m}$, a much smaller value than the minimal average urban block size (about $10\textrm{m}\times10\textrm{m}$).  Thus fewer elementary random built-up components are found and counted at the smallest scales compared to the number that would be  expected for an ideal self-similar structure.
\par 
Image  N45-024 (Turin) is shown in  Fig.~\ref{fig:WV2_TO} (top panel).  Log-log results of $\sigma^2_{DMA}$ are plotted for each sub-image \textit{A, B, C, D} for the whole range of $s$ scales (bottom panels). The slope is estimated by ordinary linear regression over three different ranges of $s$ values. $H_{1}$, ${H}$  and  $H_{2}$ corresponding respectively to the first, intermediate and last decades of $s$ are reported in Table \ref{tab:WV2_Results}.  $H_{2}$ ranges between $0.10 \div  0.15$, $H$ ranges between $0.12 \div  0.16$,   while $H_{1}$ ranges between  $0.23 \div  0.24$. The Hurst exponent of section \textit{D} is the highest and indeed corresponds to less urbanised areas (Torino hills). In Table \ref{tab:WV2_Results} further results are reported for other Turin areas (image N45-037 and  N45-124).
\par
Image N  48-181 (Vienna) is shown in Fig.~\ref{fig:WV2_WI} (top panel) and 
 the $\sigma_{DMA}^2$ results are plotted in log-log scale (bottom panels). Sections \textit{A}, \textit{B} and \textit{C} are highly urbanized areas, while Section \textit{D} is less urbanized. This is reflected in the Hurst exponent estimates, which tends to be lower for urbanized areas. 
 $H_{2}$ ranges between $0.09 \div  0.17$,  $H$ ranges between $0.12 \div  0.20$, while  $H_{1}$ ranges between $0.22 \div 0.27$ (Table \ref{tab:WV2_Results}).  In Table \ref{tab:WV2_Results} further results are reported for other areas of Vienna (image N48-006 and  N48-465).
\par 
Image  N 47-377 (Zurich) is shown in Fig.\ref{fig:WV2_ZU} (top panel) and 
 the $\sigma_{DMA}^2$ results are plotted in log-log scale (bottom panels). The most densely urbanized area looks Section \textit{B}, while the least Section \textit{A}; overall, the city of Zurich seems more heterogeneous compared to Turin (Fig.~\ref{fig:WV2_TO}) and Vienna (Fig.~\ref{fig:WV2_WI}). Large wooded areas interrupt frequently the urbanized grid. This is reflected in the Hurst exponent, which takes higher and less diversified values than for Torino and Vienna. 
  $H_{2}$ ranges between $0.10 \div  0.20$, $H$ ranges between $0.12 \div  0.22$, while  $H_{1}$ ranges between $0.22 \div 0.28$   (Table \ref{tab:WV2_Results}). 
  Further results are reported for other areas of Zurich (images N47-167 and  N48-230) in Table \ref{tab:WV2_Results}.
\par \medskip
The image  N50-090 of the city of Prague is shown in  Fig.~\ref{fig:WV2_PR} (top panel) and  
 $\sigma_{DMA}^2$ results are reported in log-log scale (bottom panels) for each of the four sections of the whole image.   $H_{2}$ ranges between $0.11 \div  0.16$, $H$ ranges between $0.14 \div  0.17$, while  $H_{1}$ takes the value  $0.26$.  
  Further results are reported for other areas of Prague (images N50-045 and  N49-908) in Table \ref{tab:WV2_Results}.

\begin{table*}[ht]
\centering
\subfloat{
\begin{tabular}{cc|cccc|c} 
  Torino &  & $H_{1}$ & $H$ & $H_{2}$ & $R^2$ & $D_f$\\ 
  \hline \multirow{4}{*}{N 45-024} 
 & A & 0.23 & 0.12 & 0.10 & 0.93 & 1.90 \\ 
 & B & 0.23 & 0.13 & 0.11 & 0.95 & 1.89 \\ 
 & C & 0.24 & 0.13 & 0.11 & 0.94 & 1.89 \\ 
 & D & 0.25 & 0.16 & 0.15 & 0.97 & 1.85 \\ 
  \hline \multirow{4}{*}{N 45-037} 
 & A & 0.31 & 0.30 & 0.32 & 1.00 & 1.68 \\ 
 & B & 0.27 & 0.25 & 0.27 & 0.99 & 1.73 \\ 
 & C & 0.28 & 0.23 & 0.23 & 0.99 & 1.77 \\ 
 & D & 0.33 & 0.30 & 0.31 & 1.00 & 1.69 \\ 
  \hline \multirow{4}{*}{N 45-124} 
 & A & 0.30 & 0.28 & 0.30 & 1.00 & 1.70 \\ 
 & B & 0.27 & 0.22 & 0.23 & 0.99 & 1.77 \\ 
 & C & 0.25 & 0.16 & 0.13 & 0.97 & 1.87 \\ 
 & D & 0.26 & 0.17 & 0.15 & 0.98 & 1.85 \\ 
\end{tabular}
}
\qquad
\subfloat{
\begin{tabular}{cc|cccc|c}
        Vienna &  & $H_{1}$ & ${H}$ & $H_{2}$ & $R^2$ & $D_f$\\
        \hline
        \multirow{4}{*}{N 48-181}   & A & 0.23  & 0.13 & 0.11 & 0.94 & 1.89\\
                                           & B & 0.24  & 0.15 & 0.13 & 0.97 & 1.87\\
                                        & C & 0.22  & 0.12 & 0.09 & 0.93 & 1.90\\
                                        & D & 0.27  & 0.20 & 0.17 &  0.98 & 1.83\\
        \hline
        \multirow{4}{*}{N 48-006}   & A & 0.33  & 0.26 & 0.23  & 0.99 & 1.77\\
                                        & B &0.38  & 0.30 & 0.27 & 0.99 & 1.73\\
                                        & C & 0.33   & 0.29 & 0.27 & 1.00 & 1.73\\
                                        & D & 0.33  & 0.26 & 0.22 & 0.99 & 1.78\\
        \hline
        \multirow{4}{*}{N 48-465}   & A &  0.41  & 0.30 & 0.27 & 0.99 & 1.73\\
                                        & B & 0.39  & 0.32 & 0.31  & 1.00 & 1.69\\
                                        & C & 0.40  & 0.29 & 0.26  & 0.99 & 1.74\\
                                        & D & 0.40  & 0.28 & 0.24 & 0.98 & 1.76\\
    \end{tabular}
    }
\\
\subfloat{
\begin{tabular}{cc|cccc|c} 
  Zurich &  & $H_{1}$ & $H$ & $H_{2}$ & $R^2$ & $D_f$\\ 
  \hline \multirow{4}{*}{N 47-377} 
 & A & 0.28 & 0.22 & 0.20 & 0.99 & 1.80 \\ 
 & B & 0.27 & 0.19 & 0.17 & 0.98 & 1.83 \\ 
 & C & 0.23 & 0.12 & 0.10 & 0.93 & 1.90 \\ 
 & D & 0.22 & 0.14 & 0.12 & 0.96 & 1.88 \\ 
  \hline \multirow{4}{*}{N 47-167} 
 & A & 0.26 & 0.19 & 0.16 & 0.98 & 1.84 \\ 
 & B & 0.27 & 0.20 & 0.19 & 0.99 & 1.81 \\ 
 & C & 0.28 & 0.23 & 0.23 & 0.99 & 1.77 \\ 
 & D & 0.26 & 0.18 & 0.16 & 0.98 & 1.84 \\ 
  \hline \multirow{4}{*}{N 47-230} 
 & A & 0.28 & 0.23 & 0.21 & 0.99 & 1.79 \\ 
 & B & 0.27 & 0.21 & 0.19 & 0.99 & 1.81 \\ 
 & C & 0.29 & 0.23 & 0.21 & 0.99 & 1.79 \\ 
 & D & 0.27 & 0.21 & 0.18 & 0.99 & 1.82 \\ 
\end{tabular} }
\qquad
\subfloat{
\begin{tabular}{cc|cccc|c}
        Prague  &  & $H_{1}$ & ${H}$ & $H_{2}$ & $R^2$ & $D_f$\\
        \hline
        \multirow{4}{*}{N 50-090} & A & 0.26  & 0.17 & 0.16 & 0.97 & 1.84\\
                                        & B & 0.26  & 0.15 & 0.12 &  0.95 & 1.88\\
                                        & C & 0.26  & 0.16 & 0.14 & 0.96 & 1.86\\
                                        & D &0.26   & 0.14 & 0.11 & 0.93 & 1.89\\
        \hline
        \multirow{4}{*}{N 50-045} & A & 0.25  & 0.18 & 0.17  & 0.99 & 1.83\\
                                        & B & 0.24  & 0.14 & 0.12 & 0.95 & 1.88\\
                                        & C & 0.25  & 0.19 & 0.19 & 0.99 & 1.81\\
                                        & D &0.26   & 0.21 & 0.21 & 0.99 & 1.79\\
        \hline
        \multirow{4}{*}{N 49-908} & A & 0.29  & 0.24 & 0.24 & 0.99 & 1.76\\
                                        & B & 0.34  & 0.32 & 0.32 & 1.00 & 1.68\\
                                        & C & 0.36   & 0.32 &0.33 & 1.00 & 1.67\\
                                        & D & 0.33  & 0.29 & 0.30 & 1.00 & 1.70\\
    \end{tabular}}
    \caption{Hurst exponents estimated for the WorldView-2 satellite images  45-024, 45-037, 45-124 (Torino);  48-181, 48-006, 48-465 (Vienna);  47-377, 47-230, 47-167 (Zurich); 50-090, 50-045, 49-908  (Prague).
    The Hurst exponents $H_{1}$  $H$ and $H_{2}$ have been obtained by implementing the $2d$-Detrending Moving Average algorithm over the first, whole and last range of decades as summarised in Section \ref{sec:methods}. For each image the Hurst exponent is estimated for 4 cross-sections (different urban areas) labelled A, B, C, D as shown  in Fig. \ref{fig:WV2_TO} for the image 45-024.  Last column reports the estimates of the fractal dimension by using $D_f=d-H$ with the Hurst exponents $H_{2}$ and $d = 2$. Using the Hurst exponents results in the second column, referred to as $H$, alternative but similar values of $D_f$ can be obtained.}
    \label{tab:WV2_Results}
\end{table*}
\par
The fractal dimension $D_f$ is reported in the last column of Tables \ref{tab:WV2_Results} for the sector A,B,C,D of the above described images. $D_f$ values have been calculated by introducing the value $H_2$ in the relationship Eq. \ref{eq:Dimension}. $D_f$ values with $H$ and $H_1$ can be easily obtained as well.

\begin{figure}[!h]
\subfloat{ 
     \begin{tabular}{l|cccc} 
        Ref. & {\cite{bettencourt2007growth}}  & {\cite{bettencourt2013origins}} & {\cite{ribeiro2017model}} & { \cite{molinero2021geometry}} \\
        \hline
        \hline
        \smallskip
               ${\beta_i}$ & $[0.74,0.92]\quad$ & $1-\frac{D_f}{d(d+D_f)} \quad$ &  $ \frac{\gamma}{D_{{f}}}$ & $\frac{D_{f}}{D_{p}}$\\ 
        \smallskip
        ${\beta_s}$  & $[1.01,1.33]\quad$ & $1+\frac{D_f}{d(d+D_f)}\quad $&  $2-\frac{\gamma}{D_{{f}}}$ & $2-\frac{D_{f}}{D_{p}}$ \\
        \hline 
    \end{tabular}
    }
   \qquad 
    \subfloat{ 
    \includegraphics[width=\linewidth]{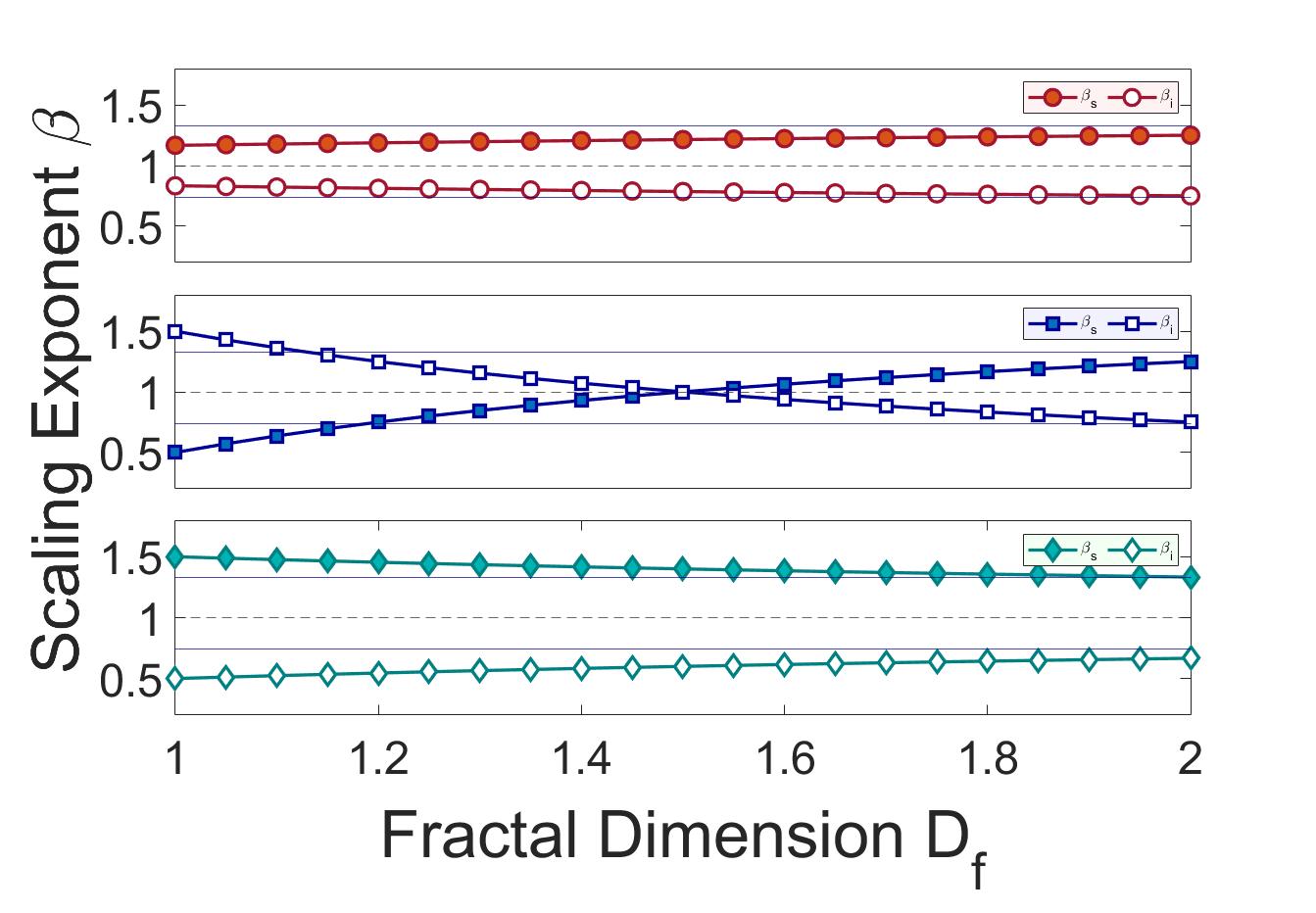}}
    \caption{Range of the empirical values for  ${\beta_i}$ and  ${\beta_s}$ reported in Ref. \cite{bettencourt2007growth} (second column). Analytical expression of the exponents ${\beta_i}$ and  ${\beta_s}$ deduced in Refs. \cite{bettencourt2013origins,ribeiro2017model,molinero2021geometry} (respectively third, fourth and fifth columns). Plot of $\beta_{s}$ (filled symbols) and  $\beta_{i}$ (hollow symbols) as defined by Eqs.~(\ref{eq:betaisBettencourt}) (red circle), Eqs.~(\ref{eq:betaisRibeiro}) (blue square) Eqs.~(\ref{eq:betaisMolinero}) (green diamond). A different dependence of the exponents on the fractal dimension is observed: $\beta_{s}$ increases ($\beta_{i}$ decreases) very slowly with the fractal dimension $D_f$ according to  Ref. \cite{bettencourt2013origins}; a stronger increase (decrease) is found according to   Ref. \cite{ribeiro2017model};  whereas   $\beta_{s}$ decreases ($\beta_{i}$ increases) with $D_f$ according to  Ref.\cite{molinero2021geometry}.}
    \label{fig:BRM_theory}
\end{figure}

\section{DISCUSSION}
\label{sec:discussion}
In this section, the values of $H$ and $D_f$ obtained in Section \ref{sec:data} and summarized in Table~\ref{tab:WV2_Results}, will be compared  with those   reported in previous works \cite{batty1987fractal,frankhauser1998fractal,shen2002fractal,encarnaccao2012fractal,tannier2013defining, chen2013set,emerson2005comparison,liang2013evaluation,liang2018characterizing}. 
Furthermore, the urban scaling exponents $\beta$   will be estimated by introducing  the numerical results of $D_f$  into the relationships linking $\beta$ and $D_f$ worked out in \cite{bettencourt2013origins,ribeiro2017model,molinero2021geometry}. 
For the sake of the discussion, a  brief summary  of previous studies dealing with fractal measurements and scaling models of urban areas is reported here below. 
\par 
The fractal dimension $D_f$  of cities has been estimated by approaches as diverse as box-counting, radial method  \cite{shen2002fractal,encarnaccao2012fractal,chen2013set,tannier2013defining} isarithm, variogram   \cite{emerson2005comparison,liang2013evaluation,liang2018characterizing}.
With embedding dimension  $d=2$ and $d=3$, the  fractal dimension $D_f$  takes value respectively in  the range $1.0 \div 2.0$ and $2.0 \div 3.0$. The Hurst exponent takes a unique value regardless of $d$ allowing the comparison of results obtained by different methods.
Fractal dimensions ranging  between $1.28 \leq D_f \leq 1.70$ have been reported  for Omaha and New York City  in \cite{shen2002fractal},     between $1.44 \leq D_f \leq 1.62$,  and $ 1.68 \leq D_f \leq 1.50$, for  Belgium’s 18 largest cities in \cite{tannier2013defining}.  Values  in the range $1 \leq D_f < 1.26 $, for dispersed areas, $1.26 \leq D_f < 1.54$ for new seeds of urbanised areas, $1.54 \leq D_f \leq 2$ for densely urbanized and consolidated areas are reported for Lisbon in Ref.~\cite{encarnaccao2012fractal}. 
Several mega-cities and mining cities of China  are investigated over different periods: the fractal dimension ranges between $1.57 \leq D_f \leq 1.74$ in 1990, and $1.57 \leq D_f \leq 1.78$ in 2000  \cite{chen2013set} and between $1.62 \leq D_f \leq 1.80$ \cite{emerson2005comparison}. 
Fractal dimensions of satellite images of cities are obtained by (i) isarithm,  (ii) triangular prism and  (iii) variogram   ranging respectively between (i) $2.80 \leq D_f \leq 3.00$; (ii) $2.60 \leq D_f \leq 2.80$, for urban, forest and grass, $2.30 \leq D_f \leq 2.80$ for cropland and pasture; $2.20 \leq D_f \leq 2.60$ for water; (iii) $2.80 \leq D_f \leq 3.00$ for cropland and water; $D_f \geq 3.00$ for urban, forest and grass  \cite{liang2013evaluation}.
The triangular prism  yields  lower  values of $D_f$ compared to those obtained by isarithm and variogram methods. The images analysed by triangular prism date back to 1975, while the other images were acquired in 2000 (isarithm and variogram). After 25 years, the city had become a large metropolis where roads, highways, and buildings filled the area. Such changes in the urban landscape  can reasonably explain the increased value of $D_f$ and the corresponding decrease of $H$.
Fractal dimension of red band satellite images of the Indianapolis area ranges respectively between $2.72 \leq D_f \leq 2.82$  (isarithm), $2.78 \leq D_f \leq 2.93$  (triangular prism), and $2.88 \leq D_f \leq 2.96$ (variogram)  Ref.~\cite{liang2018characterizing}.
On average the $H$ values obtained  by using  satellite images are smaller than those obtained by using traditional data sets as cartographic maps. 
\par 
To further substantiate our study,  the values of the Hurst exponent $H$ and of the fractal dimension $D_f$ of the satellite images as those  in Figs.~\ref{fig:WV2_TO}-\ref{fig:WV2_PR} will be validated against the the relationships linking $\beta$ and $D_f$ deduced in  Refs. \cite{bettencourt2013origins,ribeiro2017model,molinero2021geometry} briefly recalled below.
\par
Under the assumption of incremental network growth and bounded human effort, infrastructural and the socio-economic features  are written in Ref. \cite{bettencourt2013origins} as power laws of the population size  respectively  with exponents:
\begin{equation}
    \beta_i=1-\frac{D_f}{d(d+D_f)} \,  \quad \quad
    \beta_s=1+\frac{D_f}{d(d+D_f)} \, 
    \label{eq:betaisBettencourt}
\end{equation}
In Ref.~\cite{ribeiro2017model}, the interaction strength between individuals is modelled in terms of a scalar field  varying inversely with the distance and the total  interaction intensity is obtained in the form of a power law of the population size, with scaling exponents respectively for the  infrastructural and socio-economic quantities:
\begin{equation}
    \beta_{{i}} = \frac{\gamma}{D_{{f}}} \, \quad \quad \quad \beta_{{s}} = 2-\frac{\gamma}{D_{{f}}} \,  \quad ,
    \label{eq:betaisRibeiro}
\end{equation} 
with  $\gamma$ varying in the range $1.0\div 1.5$ and $\gamma=1.0$ corresponds to the Newtonian gravitational law in $d=2$.
In the long-range interaction regime $\gamma / D_{{f}}<1$,  $\beta_{{s}}>1$  implying that  superlinear  socio-economic scaling behaviour occurs when the individuals  can interact with all other individuals of the city. 
In Ref.~\cite{molinero2021geometry} a three-dimensional building infrastructure is considered with the  socio-economic interactions occurring in a $3D$ fractal cloud. The scaling exponents for the infrastructures and the socio-economic activities are written respectively:
\begin{equation}
    \beta_{i}=\frac{D_{f}}{D_{p}}\, \quad \quad \quad\beta_{s}= 2-\frac{D_{f}}{D_{p}} \,\quad .
    \label{eq:betaisMolinero}
\end{equation}
For the ease of the discussion, Eqs.~(\ref{eq:betaisBettencourt}-\ref{eq:betaisMolinero}) are gathered together in the top panel of  Fig. \ref{fig:BRM_theory} (third, fourth and fifth columns) with the empirical values of $\beta_{i}$ and  $\beta_{s}$ (second column) reported in \cite{bettencourt2007growth}. 
The derivatives of the scaling exponents $\beta_{i}$ and $\beta_{s}$ with respect to $D_f$ yield  respectively: $\partial \beta_s/ \partial D_f= 1/(d + D_f)^2 \,$ ;
$\partial \beta_s/ \partial D_f= \gamma/ D_f^2 \,$;
 $ \partial \beta_s/ \partial D_f= -1/(1 + D_f)^2 \, $ (the derivatives $\partial \beta_i/ \partial D_f$ are the same but with opposite sign).
The exponents $\beta_{i}$ and $\beta_{s}$  exhibit a different dependence on $D_f$, as  Fig.~\ref{fig:BRM_theory} also shows. In particular, the exponent $\beta_{s}$ increases ($\beta_{i}$ decreases) very slowly with the fractal dimension $D_f$ according to Eqs.~(\ref{eq:betaisBettencourt}). A steeper increase of  $\beta_{s}$ (decrease of $\beta_{i}$) is found according to the Eqs.~(\ref{eq:betaisRibeiro}) which exhibit an interesting behaviour:  at $D_f=1.5$,  $\beta_{i}$ and $\beta_{s}$  become respectively larger and smaller than 1. The inversion can be related to the different constraints posed by a urban area mostly distributed along a one-dimensional geometrical structure, with fractal dimension $D_f \rightarrow 1$.  Such topological constraint implies that the cost of the physical infrastructure exceeds the socioeconomic urban organization advantage. $D_f \rightarrow 2$ corresponds to a compact urban structure almost regularly distributed over a two-dimensional surface, where the cost of the infrastructure are fully compensated by the socio-economic development advantage.
Surprisingly.   $\beta_{s}$ decreases ($\beta_{i}$ increases) with $D_f$ according to Eqs.~(\ref{eq:betaisMolinero}) and  Ref. \cite{molinero2021geometry}. In this case, the behaviour of $\beta_{s}$  does not exhibit the increasing dependence on $D_f$  expected on account of previous studies.

\begin{table*}[]
  \centering
       \begin{tabular}{c|c|cccc|cccc|cccc}
   
  \multicolumn{2}{c|}{Ref.} &  \multicolumn{4}{c|}{ \cite{bettencourt2013origins}} &
      \multicolumn{4}{c|}{ \cite{ribeiro2017model}}& \multicolumn{4}{c}{ \cite{molinero2021geometry}}\\
       \hline \hline
         \multicolumn{2}{c|}{Vienna}  & \multicolumn{1}{c}{A} & \multicolumn{1}{c}{B} & \multicolumn{1}{c}{C} & \multicolumn{1}{c|}{D} &
        \multicolumn{1}{c}{A} & \multicolumn{1}{c}{B} & \multicolumn{1}{c}{C} & \multicolumn{1}{c|}{D} &
         \multicolumn{1}{c}{A} & \multicolumn{1}{c}{B} & \multicolumn{1}{c}{C} & \multicolumn{1}{c}{D}\\
        \cline{3-14}
        \hline
        \hline
        \multirow{2}{*}{N48-181}
        & $\beta_i$ & 0.76 & 0.76 & 0.76 & 0.76 & 0.80 & 0.81 & 0.80 & 0.83 & 0.65 & 0.65 & 0.65 & 0.64\\
                      & $\beta_s$ & 1.24 & 1.24 & 1.24 & 1.24 & 1.20 & 1.20 & 1.20 & 1.17& 1.35 & 1.35 & 1.35 & 1.36\\
                             \hline
        \multirow{2}{*}{N48-006}
        & $\beta_i$ &  0.76  &  0.77   & 0.77  &  0.76 &  0.85  &  0.87 &   0.87  &  0.84 & 0.64 &   0.63   & 0.63 &  0.64\\
                      & $\beta_s$ &  1.23  &  1.23  &  1.23 &   1.23 & 1.15 &   1.13  &  1.13  &  1.16 & 1.36  &  1.37 &   1.37  &  1.36\\
                             \hline
                            \multirow{2}{*}{N 48-465}
        & $\beta_i$ & 0.77 &   0.77   &  0.77  &  0.77 &   0.87  &  0.89  &   0.86  &  0.85& 0.63 &   0.63  &  0.63  &  0.64\\
                      & $\beta_s$ &    1.23  &   1.23  &  1.23  &  1.23& 1.13  &   1.11 &  1.14  &  1.15 & 1.37  &  1.37   &  1.36  &  1.36\\
\hline  \hline
        \multicolumn{2}{c|}{ Prague} & \multicolumn{1}{c}{A} & \multicolumn{1}{c}{B} & \multicolumn{1}{c}{C} & \multicolumn{1}{c|}{D} &
        \multicolumn{1}{c}{A} & \multicolumn{1}{c}{B} & \multicolumn{1}{c}{C} & \multicolumn{1}{c|}{D} &
         \multicolumn{1}{c}{A} & \multicolumn{1}{c}{B} & \multicolumn{1}{c}{C} & \multicolumn{1}{c}{D}\\
        \cline{3-14}
        \hline
          \hline
        \multirow{2}{*}{ N 50-090} 
        & $\beta_i$ & 0.76 & 0.76 & 0.76 & 0.76& 0.82 & 0.81 & 0.81 & 0.80 & 0.65 & 0.65 & 0.65 & 0.65\\
        & $\beta_s$ & 1.24 & 1.24 & 1.24 & 1.24& 1.18 & 1.19 & 1.18 & 1.19& 1.35 & 1.35 & 1.35 & 1.35\\
                \hline
        \multirow{2}{*}{N 50-045} 
        & $\beta_i$ & 0.76  &   0.76 &   0.76  &  0.76 & 0.82  &  0.80  &  0.83  &  0.84 & 0.65 &   0.65 &   0.64  &  0.64\\
        & $\beta_s$ &  1.24  &  1.24  & 1.24  &  1.24 & 1.18  &  1.20 &   1.17  &  1.16 & 1.35  &  1.35  &  1.36 &   1.36\\ 
              \hline
        \multirow{2}{*}{ N 50-908} 
        & $\beta_i$ & 0.77  &  0.77  &  0.77 &   0.77 & 0.85  &  0.89 &  0.89 &   0.88 & 0.64  &  0.63 &   0.62  &  0.63
\\
        & $\beta_s$ & 1.23   & 1.23   & 1.23  & 1.23 & 1.15  &  1.11 &   1.10 &   1.12 & 1.36  &  1.37 &   1.37  &  1.37\\  
         \hline  \hline 
         \multicolumn{2}{c|}{Torino}  & \multicolumn{1}{c}{A} & \multicolumn{1}{c}{B} & \multicolumn{1}{c}{C} & \multicolumn{1}{c|}{D} &
        \multicolumn{1}{c}{A} & \multicolumn{1}{c}{B} & \multicolumn{1}{c}{C} & \multicolumn{1}{c|}{D} &
         \multicolumn{1}{c}{A} & \multicolumn{1}{c}{B} & \multicolumn{1}{c}{C} & \multicolumn{1}{c}{D}\\
        \cline{3-14}
        \hline
          \hline
        \multirow{2}{*}{N 45-024} 
        & $\beta_i$ &   0.76 &   0.76  &   0.76  &  0.76  &  0.79 &   0.79 &   0.79   & 0.81 & 0.65 &   0.65  &  0.65  &  0.65\\
        & $\beta_s$ & 1.24  &   1.24   &  1.24 &    1.24&  1.21 &   1.21  &  1.21  &  1.19 &  1.34  &  1.35  & 1.35   & 1.35 \\
                \hline
        \multirow{2}{*}{N 45-037} 
        & $\beta_i$ & 0.77 &    0.77 &    0.76  &   0.77  &    0.89 &   0.87  &  0.85  &  0.89  & 0.63  &  0.63  &   0.64   & 0.63 \\
        & $\beta_s$ &  1.23  &   1.23 &    1.23 &    1.23 & 1.11   &  1.13  &   1.15  &   1.11 & 1.37 &    1.37 &   1.36  &  1.37 \\ 
        \hline
        \multirow{2}{*}{N 45-124} 
        & $\beta_i$ &  0.77   &  0.76   &  0.76  &   0.76 &  0.88 &   0.85 &   0.80  &  0.81 &  0.63 &   0.64  &  0.65 &   0.65
        \\
        & $\beta_s$ & 1.23 &    1.23   &  1.24 &    1.24 &  1.12  &  1.15  &  1.20 &   1.19 & 1.37   &  1.36  &  1.35   & 1.35\\  
        \hline  \hline
        \multicolumn{2}{c|}{Zurich}  & \multicolumn{1}{c}{A} & \multicolumn{1}{c}{B} & \multicolumn{1}{c}{C} & \multicolumn{1}{c|}{D} &
        \multicolumn{1}{c}{A} & \multicolumn{1}{c}{B} & \multicolumn{1}{c}{C} & \multicolumn{1}{c|}{D} &
        \multicolumn{1}{c}{A} & \multicolumn{1}{c}{B} & \multicolumn{1}{c}{C} & \multicolumn{1}{c}{D}\\
        \cline{3-14}
        \hline
          \hline
        \multirow{2}{*}{N 47-377} 
        & $\beta_i$ &    0.76 &    0.76  &   0.76&     0.76 & 0.83   &  0.82 &    0.79& 0.80 & 0.64 &    0.65&     0.65  &   0.65 \\
        & $\beta_s$ &  1.24  &  1.24  &  1.24   & 1.24 &
        1.17 &   1.18  &  1.21  &  1.20 & 
        1.36   & 1.35  &  1.34  &  1.35     \\
        \hline
        \multirow{2}{*}{N 47-167} 
        & $\beta_i$ &0.76  &   0.76&    0.76 &    0.76&  0.81 &   0.83 &    0.85  &   0.81 & 0.65 &    0.64  &   0.64 &    0.65\\
        & $\beta_s$ &  1.24 &   1.24 &   1.23 &   1.24  &
        1.18 &   1.17 &   1.15 &    1.18 &
        1.35 &  1.36 &   1.36  &  1.35 \\ 
        \hline
        \multirow{2}{*}{N 47-230} 
        & $\beta_i$ & 0.76 &    0.76  &   0.76 &    0.76 & 
        0.84 &    0.83 &    0.84  &   0.82 & 
        0.64 &    0.64 &    0.64&     0.64  \\
        & $\beta_s$ & 1.24   & 1.24  &  1.24  &  1.24 &
        1.16 &   1.17  &  1.16   & 1.18&
        1.36 &   1.36 &   1.36  &  1.35 \\  
    \end{tabular}
    \caption{
    Scaling exponents $\beta_i$ and $\beta_s$ obtained  by using the values of $D_f$ reported in Table \ref{tab:WV2_Results}. The values for the sections A, B, C, D are obtained by using  Eqs. (\ref{eq:betaisBettencourt}), Eqs. (\ref{eq:betaisRibeiro})  with $\gamma = 1.5$ and Eqs. (\ref{eq:betaisMolinero}), with $D_p = D_f + 1$.
    }
    \label{tab:brm_exp}
\end{table*}

\begin{figure*}[]
    \includegraphics[width=0.69\linewidth]{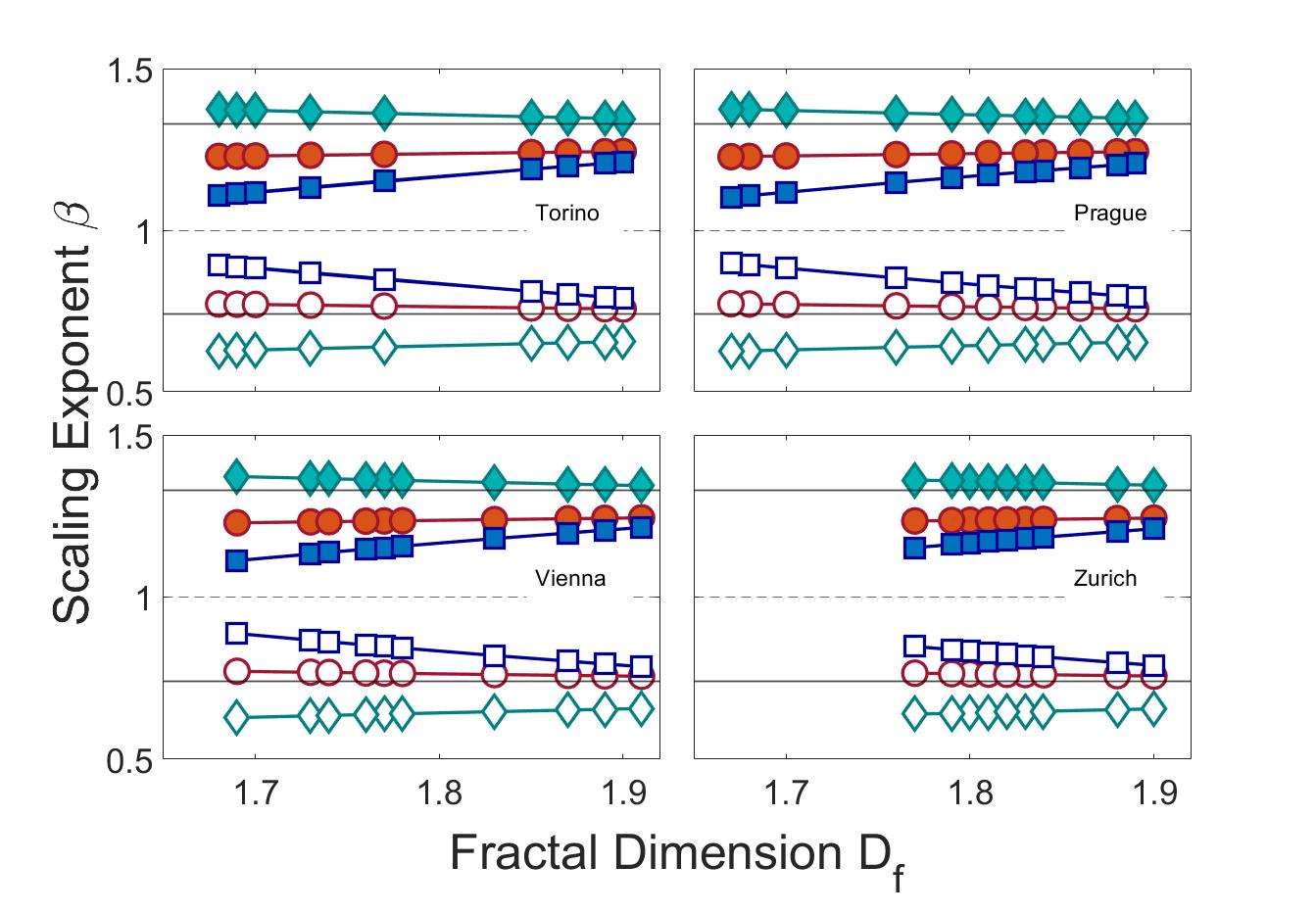}%
    \caption{Experimental exponents $\beta_{s}$ (filled symbols) and  $\beta_{i}$ (hollow symbols) as defined by Eqs.~(\ref{eq:betaisBettencourt}) (circle), Eqs.~(\ref{eq:betaisRibeiro}) (square) and Eqs.~(\ref{eq:betaisMolinero}) (diamond) for different areas of Torino, Zurich, Vienna and  Prague. A different dependence of the exponents on the fractal dimension is observed.}
    \label{fig:BRM_data}
\end{figure*}

\par 
The fractal dimension values (ranging between $1.6 \leq D_f \leq 1.8$ and $1 < D_f < 2$) used for the scaling law estimates were taken from third party sources in \cite{bettencourt2013origins,ribeiro2017model,molinero2021geometry}.
In this work, the exponents $\beta_{i}$ and $\beta_{s}$ are calculated by introducing the values of $D_f$ (Table \ref{tab:WV2_Results}) into the  Eqs.~(\ref{eq:betaisBettencourt}-\ref{eq:betaisMolinero}).  Values  are shown in Table \ref{tab:brm_exp}.  Columns from 3 to 6 show the values obtained by Eqs.~(\ref{eq:betaisBettencourt});  Eqs. (\ref{eq:betaisRibeiro}) with $\gamma = 1.5$  correspond to columns from 7 to 10); Eqs. (\ref{eq:betaisMolinero}) with $D_p = D_f + 1$ correspond to columns from 11 to 14.
The analyzed areas have  infrastructures scaling sub-linearly and socio-economic interactions scaling super-linearly with exponents in the range of empirical values according to Ref.\cite{bettencourt2007growth} when  Eqs.~(\ref{eq:betaisBettencourt}) and Eqs. (\ref{eq:betaisRibeiro}) are used.
The values of the exponent yielded by Eqs. (\ref{eq:betaisMolinero}) systematically exceed the expected values.  
The values are plotted in Fig. \ref{fig:BRM_data}, where the range of empirical values (column 2 of the Table in \ref{fig:BRM_theory})  are also  indicated by  thin horizontal lines.
\par \bigskip

\section{Conclusions}
\label{sec:conclusions}
\par \medskip
This work enriches the existing literature on two fronts. First, it provides a new method for urban classification capable of distinguishing different areas such as urban and  suburban areas. In particular, the Hurst exponent $H$ (resp., the fractal dimension $D_f$) is smaller (larger) for highly urbanized areas and larger (smaller) for detached rural areas. The Hurst exponent $H$  of several large European cities has been estimated  by implementing the Detrended Moving Average algorithm on high resolution remotely sensed images (WorldView-2 Urban Atlas database). The values of $H$ are linked to the fractal dimension $D_f$ through the relationship (\ref{eq:Dimension}).  
Our estimates provide   $0.10 \leq H \leq 0.30$ for the Hurst exponent, which correspond to fractal dimensions ranging between $1.65 \leq D_f \leq 1.90$. 
Interestingly,  we obtain slightly smaller Hurst exponent and higher fractal dimension on average with respect to the estimates of the urban fractal dimensions reported in  \cite{shen2002fractal,encarnaccao2012fractal,tannier2013defining,chen2013set}. 
Our values of the Hurst exponent are closer to those provided in Refs.~\cite{emerson2005comparison,liang2013evaluation,liang2018characterizing}.
This result seems to suggest that highly reproducible values are obtained when satellite images are used as opposed to those provided by other data sets.  
\par 
Second, the manuscript demonstrates that a geometrical approach to urban scaling theory, which exploit the statistical structure of high resolution satellite images of cities, provides robust estimates and validation of urban scaling laws. 
A rich theory has developed a number of models that describe the characteristic power law behavior of features exhibiting super-linear or sub-linear scaling respectively for socio-economic and infrastructural variables. 
Interestingly, for the quantification of such formulae, the theoretical framework relies on fractal measures. 
By using the definitions of the scaling exponents reported in the table at the top of Figure \ref{fig:BRM_theory}, $\beta_i$ and $\beta_s$ can be calculated. 
The results for the  images N45-024,  N48-181, N47-377 and N50-090 of the cities of Turin, Vienna, Zurich and Prague are reported in Table \ref{tab:brm_exp} and plotted in Figure \ref{fig:BRM_data}.
Thus, the proposed method can be used alone or in combination with other measures and approaches to provide significant new insights in urban scaling model analysis and in designing the related needs for intervention and policy-making activities. 

\bigskip
\section*{Acknowledgments}  
\par
This work received financial support from the FuturICT2.0 project (a FLAG-ERA Initiative within the Joint Transnational Calls 2016, Grant Number: JTC-2016-004) and from the SIP project (Italian Ministry of Economic Development (MISE) Programme  on  "Emerging Technologies in the context of 5G").


\begin{thebibliography}{10}

\bibitem{gao2019computational}
Jian Gao, Yi-Cheng Zhang, and Tao Zhou.
\newblock Computational socioeconomics.
\newblock {\em Physics Reports}, 817:1--104, 2019.

\bibitem{bettencourt2007growth}
Lu{\'\i}s~MA Bettencourt, Jos{\'e} Lobo, Dirk Helbing, Christian K{\"u}hnert,
  and Geoffrey~B West.
\newblock Growth, innovation, scaling, and the pace of life in cities.
\newblock {\em Proceedings of the national academy of sciences},
  104(17):7301--7306, 2007.

\bibitem{bettencourt2013origins}
Lu{\'\i}s~MA Bettencourt.
\newblock The origins of scaling in cities.
\newblock {\em Science}, 340(6139):1438--1441, 2013.

\bibitem{ribeiro2017model}
Fabiano~L Ribeiro, Joao Meirelles, Fernando~F Ferreira, and Camilo~Rodrigues
  Neto.
\newblock A model of urban scaling laws based on distance dependent
  interactions.
\newblock {\em Royal Society Open Science}, 4(3):160926, 2017.

\bibitem{molinero2021geometry}
Carlos Molinero and Stefan Thurner.
\newblock How the geometry of cities determines urban scaling laws.
\newblock {\em Journal of the Royal Society Interface}, 18(176):20200705, 2021.

\bibitem{portugali2012self}
Juval Portugali.
\newblock {\em Self-organization and the city}.
\newblock Springer Science \& Business Media, 2012.

\bibitem{haken2021urban}
Hermann Haken and Juval Portugali.
\newblock Urban scaling, urban regulatory focus and their interrelations.
\newblock {\em Synergetic Cities: Information, Steady State and Phase
  Transition: Implications to Urban Scaling, Smart Cities and Planning}, pages
  199--215, 2021.

\bibitem{mandel1}
B.B. Mandelbrot.
\newblock {\em Fractals: Form, Chance and Dimension}.
\newblock Freeman, 1977.

\bibitem{batty1987fractal}
Michael Batty and Paul~A Longley.
\newblock Fractal-based description of urban form.
\newblock {\em Environment and planning B: Planning and Design},
  14(2):123--134, 1987.

\bibitem{frankhauser1998fractal}
Pierre Frankhauser.
\newblock The fractal approach. a new tool for the spatial analysis of urban
  agglomerations.
\newblock {\em Population: an english selection}, pages 205--240, 1998.

\bibitem{shen2002fractal}
Guoqiang Shen.
\newblock Fractal dimension and fractal growth of urbanized areas.
\newblock {\em International Journal of Geographical Information Science},
  16(5):419--437, 2002.

\bibitem{tannier2013defining}
C{\'e}cile Tannier and Isabelle Thomas.
\newblock Defining and characterizing urban boundaries: A fractal analysis of
  theoretical cities and belgian cities.
\newblock {\em Computers, Environment and Urban Systems}, 41:234--248, 2013.

\bibitem{encarnaccao2012fractal}
Sara Encarna{\c{c}}{\~a}o, Marcos Gaudiano, Francisco~C Santos, Jos{\'e}~A
  Tened{\'o}rio, and Jorge~M Pacheco.
\newblock Fractal cartography of urban areas.
\newblock {\em Scientific Reports}, 2(1):1--5, 2012.

\bibitem{chen2013set}
Yanguang Chen.
\newblock A set of formulae on fractal dimension relations and its application
  to urban form.
\newblock {\em Chaos, Solitons \& Fractals}, 54:150--158, 2013.

\bibitem{emerson2005comparison}
Charles~W Emerson, Nina Siu-Ngan Lam, and Dale~A Quattrochi.
\newblock A comparison of local variance, fractal dimension, and moran's i as
  aids to multispectral image classification.
\newblock {\em International Journal of Remote Sensing}, 26(8):1575--1588,
  2005.

\bibitem{liang2013evaluation}
Bingqing Liang, Qihao Weng, and Xiaohua Tong.
\newblock An evaluation of fractal characteristics of urban landscape in
  indianapolis, usa, using multi-sensor satellite images.
\newblock {\em International journal of remote sensing}, 34(3):804--823, 2013.

\bibitem{liang2018characterizing}
Bingqing Liang and Qihao Weng.
\newblock Characterizing urban landscape by using fractal-based texture
  information.
\newblock {\em Photogrammetric Engineering \& Remote Sensing}, 84(11):695--710,
  2018.

\bibitem{rozenfeld2011area}
Hern{\'a}n~D Rozenfeld, Diego Rybski, Xavier Gabaix, and Hern{\'a}n~A Makse.
\newblock The area and population of cities: New insights from a different
  perspective on cities.
\newblock {\em American Economic Review}, 101(5):2205--25, 2011.

\bibitem{levinson2012network}
David Levinson.
\newblock Network structure and city size.
\newblock {\em PloS one}, 7(1):e29721, 2012.

\bibitem{yakubo2014superlinear}
K~Yakubo, Y~Saijo, and D~Koro{\v{s}}ak.
\newblock Superlinear and sublinear urban scaling in geographical networks
  modeling cities.
\newblock {\em Physical Review E}, 90(2):022803, 2014.

\bibitem{wu2019transit}
Hao Wu, David Levinson, and Somwrita Sarkar.
\newblock How transit scaling shapes cities.
\newblock {\em Nature Sustainability}, 2(12):1142--1148, 2019.

\bibitem{keuschnigg2019urban}
Marc Keuschnigg, Selcan Mutgan, and Peter Hedstr{\"o}m.
\newblock Urban scaling and the regional divide.
\newblock {\em Science advances}, 5(1):eaav0042, 2019.

\bibitem{dong2020understanding}
Lei Dong, Zhou Huang, Jiang Zhang, and Yu~Liu.
\newblock Understanding the mesoscopic scaling patterns within cities.
\newblock {\em Scientific reports}, 10(1):1--11, 2020.

\bibitem{altmann2020spatial}
Eduardo~G Altmann.
\newblock Spatial interactions in urban scaling laws.
\newblock {\em Plos one}, 15(12):e0243390, 2020.

\bibitem{cottineau2017diverse}
Cl{\'e}mentine Cottineau, Erez Hatna, Elsa Arcaute, and Michael Batty.
\newblock Diverse cities or the systematic paradox of urban scaling laws.
\newblock {\em Computers, environment and urban systems}, 63:80--94, 2017.

\bibitem{arcaute2015constructing}
Elsa Arcaute, Erez Hatna, Peter Ferguson, Hyejin Youn, Anders Johansson, and
  Michael Batty.
\newblock Constructing cities, deconstructing scaling laws.
\newblock {\em Journal of the royal society interface}, 12(102):20140745, 2015.

\bibitem{rybski2019urban}
Diego Rybski, Elsa Arcaute, and Michael Batty.
\newblock Urban scaling laws, 2019.

\bibitem{elvidge1997relation}
Christopher~D Elvidge, Kimberley~E Baugh, Eric~A Kihn, Herbert~W Kroehl,
  Ethan~R Davis, and Chris~W Davis.
\newblock Relation between satellite observed visible-near infrared emissions,
  population, economic activity and electric power consumption.
\newblock {\em International Journal of Remote Sensing}, 18(6):1373--1379,
  1997.

\bibitem{ebener2005wealth}
Steeve Ebener, Christopher Murray, Ajay Tandon, and Christopher~C Elvidge.
\newblock From wealth to health: modelling the distribution of income per
  capita at the sub-national level using night-time light imagery.
\newblock {\em international Journal of health geographics}, 4(1):1--17, 2005.

\bibitem{donaldson2016view}
Dave Donaldson and Adam Storeygard.
\newblock The view from above: Applications of satellite data in economics.
\newblock {\em Journal of Economic Perspectives}, 30(4):171--98, 2016.

\bibitem{jean2016combining}
Neal Jean, Marshall Burke, Michael Xie, W~Matthew Davis, David~B Lobell, and
  Stefano Ermon.
\newblock Combining satellite imagery and machine learning to predict poverty.
\newblock {\em Science}, 353(6301):790--794, 2016.

\bibitem{wellmann2020remote}
Thilo Wellmann, Angela Lausch, Erik Andersson, Sonja Knapp, Chiara Cortinovis,
  Jessica Jache, Sebastian Scheuer, Peleg Kremer, Andr{\'e} Mascarenhas, Roland
  Kraemer, et~al.
\newblock Remote sensing in urban planning: Contributions towards ecologically
  sound policies?
\newblock {\em Landscape and Urban Planning}, 204:103921, 2020.

\bibitem{burke2021using}
Marshall Burke, Anne Driscoll, David~B Lobell, and Stefano Ermon.
\newblock Using satellite imagery to understand and promote sustainable
  development.
\newblock {\em Science}, 371(6535), 2021.

\bibitem{carbone2007algorithm}
Anna Carbone.
\newblock Algorithm to estimate the hurst exponent of high-dimensional
  fractals.
\newblock {\em Physical Review E}, 76(5):056703, 2007.

\bibitem{worldview}
Urban Atlas, last visit on 07/2021.
\newblock The ESA third party mission collection of the largest European urban
  areas recorded by the WorldView-2 satellite
  \url{https://tpm-ds.eo.esa.int/oads/access/collection/WorldView-2}.

\bibitem{FRACLAB}
FRACLAB, last visit on 07/2021.
\newblock We use the CLF algorithm included in the package FRACLAB downloadable
  at \url{https://project.inria.fr/fraclab/}.

\bibitem{carbone2010snow}
Anna Carbone, Bernardino~M Chiaia, Barbara Frigo, and Christian T{\"u}rk.
\newblock Snow metamorphism: A fractal approach.
\newblock {\em Physical Review E}, 82(3):036103, 2010.

\bibitem{turk2010fractal}
Christian T{\"u}rk, Anna Carbone, and Bernardino~M Chiaia.
\newblock Fractal heterogeneous media.
\newblock {\em Physical Review E}, 81(2):026706, 2010.

\bibitem{valdiviezo2014hurst}
Juan~C Valdiviezo-N, Raul Castro, Gabriel Crist{\'o}bal, and Anna Carbone.
\newblock Hurst exponent for fractal characterization of landsat images.
\newblock In {\em Remote sensing and modeling of ecosystems for sustainability
  Xi}, volume 9221, page 922103. International Society for Optics and
  Photonics, 2014.

\bibitem{arreola2021non}
Mario Arreola-Esquivel, Carina Toxqui-Quitl, Maricela Delgadillo-Herrera,
  Alfonso Padilla-Vivanco, Gabriel Ortega-Mendoza, and Anna Carbone.
\newblock Non-binary snow index for multi-component surfaces.
\newblock {\em Remote Sensing}, 13(14):2777, 2021.

\bibitem{safia2015multiband}
Abdelmounaime Safia and Dong-Chen He.
\newblock Multiband compact texture unit descriptor for intra-band and
  inter-band texture analysis.
\newblock {\em ISPRS journal of photogrammetry and remote sensing},
  105:169--185, 2015.

\end{thebibliography}
\end{document}